\documentclass{article}
\usepackage[a4paper,left=0.5in,right=0.5in,top=0.5in,bottom=0.5in]{geometry}
\usepackage{multirow}
\usepackage{authblk}
\usepackage{fullpage}
\usepackage{amsmath}
\usepackage{amssymb}
\usepackage{hyperref}
\usepackage{relsize}
\usepackage{adjustbox}
\usepackage{mathtools}
\usepackage{graphicx}
\usepackage{url}
\usepackage {hyperref}
\usepackage[table]{xcolor}

\title{An infectious diseases hazard map for India based on mobility and transportation networks}
\author{Onkar Sadekar, Mansi Budamagunta, G. J. Sreejith, Sachin Jain, and M. S. Santhanam\\
Indian Institute of Science Education and Research, Pune 411 008, India.}

\newcommand{\Deff}[0]{$D_{\text{eff}}$ }
\date{\today}
\begin{document}
\maketitle
\begin{abstract}
We propose a risk measure and construct an infectious diseases hazard map for India. Given an outbreak location, a hazard index is assigned to each city using an effective distance that depends on inter-city mobilities instead of geographical distance. We demonstrate its utility using an SIR model augmented with air, rail, and road data between top 446 cities. Simulations show that the effective distance from outbreak location reliably predicts the time of arrival of infection in other cities. The hazard index predictions compare well with the observed spread of SARS-CoV-2. The hazard map can be useful in other outbreaks also.
\end{abstract}
\section{Introduction}
As of July 2021, more than 19 Crore people -- about one in every 40 humans -- have been infected by the SARS-CoV-2, and about 39 Lakh have deceased \cite{WHO-Hopkins}. 
COVID-19 has escalated from a cluster of cases in China in late 2019 into an unprecedented global public health crisis.
In India, starting with a few cases in February 2020, the infection had spread to about 65 Lakh people in a span of about eight months when the first wave peaked. With the resurgence of the second wave in India since March 2021, the number of infections and deaths have witnessed a steep increase \cite{covid19india}.

The Spanish Flu of 1918 was one of the biggest pandemics to hit India, arriving in Bombay with the British-Indian army returning from the first world war in Europe \cite{mills}. 
The then annual report of the Sanitary Commissioner to the Government of India observed that 
``There is ample evidence during the first epidemic of the introduction of infection into a locality from another infected locality. The railway played a prominent part, as was inevitable. During the panic caused by the epidemic, the trains were filled with emigrants from infected centres, many of them being ill. The Post office also was an important agency in disseminating infection, also very largely through the Railway Postal Service. Lucknow, Lahore, Simla and other cities are said to have been infected in this manner" \cite{sani-comm}. 
Further, the report states that
``there is ample evidence to prove that infection in India during the second epidemic was carried from province to province and place to place within each province by travelers by rail, riverboats, carts and on foot". 
This mode of spread is also confirmed by other studies based on detailed data recorded then in Bombay and other provinces of British India \cite{mills}. Nearly one century after the Spanish Flu, long-distance travel is even more common. This has resulted in rapid spread of infections to remote corners of the world \cite{arrival-time-statistics, Barthelemy-network, Epidemic-networked}. It is expected that irrespective of the virus' innate capacity to infect, the spread from one geographical area to another is primarily caused by the mobility of the people \cite{Mobility-infection, Global-Travel-threats, Helbing-global-travel-risks, Barthelemy-mobility}.

The influence of transportation on the pattern of infection spread is evident in SARS-CoV-2 and earlier infectious diseases \cite{Science-Deff, Covid19-mobility, Covid19, Global-DS-Covid}.
One might identify two concurrent but distinct processes -- (a) evolution of infection within a small well-mixed geographical region (city/town), and (b) the inter-city transmission between the regions. The latter will depend crucially on the transport networks and the mobility patterns of people within the country \cite{India-Covid5, India-Covid6, SPPU-India_Covid, IITB-India_Covid}. A rather impractical limit is when transportation systems are entirely stopped leading to suppression of infection spread.
Most modeling efforts focused on prediction of caseloads in India \cite{India-Covid0, India-Covid1, India-Covid2, ICTS-India_Covid, India-Covid3, India-Covid4}, rather than geographical spreading patterns.

In this work, we propose an infectious diseases hazard map for India based on a reliable predictor of the arrival time of infections from a known outbreak location \cite{Science-Deff}.
Though first official COVID-19 cases were detected in Kerala, significant outbreaks (several hundred cases) were reported in April 2020 from Mumbai and Delhi \cite{covid19india}. Being large transport hubs, infection quickly spread into rest of India from these two cities. 
While Mumbai or Delhi could be the outbreak location now, in a general scenario, the outbreak location can be anywhere. It is natural to define hazard indices for every city/town based on  different potential outbreak locations. 

The question can be posed as follows -- in a network of $M$ cities/towns ($X_1, X_2, X_3, \cdots X_M$),  and if the outbreak location is $X_1$, can a hazard value be assigned to other cities/towns reflecting, not their geographical proximity but an ``effective proximity" incorporating mobility patterns? We discuss one solution and validate it using models incorporating extensive transportation networks data.

Note that the proposed hazard index (on which the hazard map is based) depends on the outbreak location and mobility patterns. The latter is a time-dependent factor. However, as the number of cases do not appreciably change in less than a day, and the data is made public only on a granularity of a day, we construct the hazard map assuming mobility averaged over few days to be representative for all times. For a hazard map at a subcontinental spatial scale such as India, each city/town is assumed to be well-mixed. In this work, the mobility data is applied to obtain a hazard map for 446 cities/towns with a population greater than one Lakh \cite{census}.


\section{Augmented SIR model framework}
\label{chap:model}
Our framework is based on the susceptible-infected-recovered (SIR) compartmental model augmented with connectivity information between towns and cities \cite{Science-Deff, SIR-meta1}. For a well-mixed population, the SIR model \cite{SIR-1,SIR-2} is given as 
\begin{align}
    \frac{\partial S(t)}{\partial t} &= -\alpha \frac{S(t)I(t)}{N}, \cr
    \frac{\partial I(t)}{\partial t} &= +\alpha \frac{S(t)I(t)}{N} - \beta I(t), \cr
    \frac{\partial R(t)}{\partial t} &= +\beta I(t).
    \label{eq:SIR}
\end{align}
In this, $S(t), I(t)$ and $R(t)$ denote the susceptible, infected, and recovered population respectively at time $t$. $\alpha$ and $\beta$ denote the infection and recovery rates. The total population $N=S(t)+I(t)+R(t)$ remains constant over time. However, the population in a large region like India is not well-mixed. Thus, in a network of $M$ cities/towns, Eqs.~\eqref{eq:SIR} apply inside each city/town as the population within can be assumed to be well-mixed. A small part of this population can travel between cities/towns according to
\begin{align}
\frac{\partial N_n(t)}{\partial t} &= \sum_{m=1}^M \left[ F_m^n - F_n^m  \right], \;\;\;
n,m = 1,2, \cdots M,
\label{eq:movement_kinetics}
\end{align}
where $N_n(t)$ denotes the population of $n^{\rm th}$ city/town at time $t$, and $F_n^m$ denotes the rate of people traveling from $n$ to $m$. $F_n^m$ together with the convention $F_n^n=0$ defines the traffic matrix. A city's population will be a constant if its total influx and outflux are equal. In this work the traffic matrix is inferred from limited, available real-life data, and generally they do not strictly satisfy this balance. However the time scales over which our simulations are performed are small enough that the small imbalances in fluxes do not change the city populations appreciably, accordingly,  we assume them to be constant in the rest of the work.

There is very little reliable data about infection acquired during transit. The probability of getting infected during transit is assumed to be zero. In our model, a susceptible traveler leaving city $n$ would remain so upon reaching city $m$ (similarly for infected or recovered travelers). With these assumptions, the SIR model incorporating the inter-city mobilities can be written as \cite{SIR-meta2}
\begin{align}
    \frac{\partial S_n(t)}{\partial t} &= -\alpha \frac{S_n(t)I_n(t)}{N_n} + \sum_m \Big[ \frac{F_m^n}{N_m} S_m(t) - \frac{F_n^m}{N_n} S_n(t) \Big], \cr
    \frac{\partial I_n(t)}{\partial t} &= +\alpha \frac{S_n(t)I_n(t)}{N_n} - \beta I_n(t)+ \sum_m \Big[ \frac{F_m^n}{N_m} I_m(t) - \frac{F_n^m}{N_n} I_n(t) \Big], \cr
    \frac{\partial R_n(t)}{\partial t} &= +\beta I_n(t)+ \sum_m \Big[ \frac{F_m^n}{N_m} R_m(t) - \frac{F_n^m}{N_n} R_n(t) \Big], \;\;\;n,m=1,2,\cdots M. 
\label{eq:SIR_traffic}
\end{align}
Upon adding the equations, we find that the total city population $S_n(t) + I_n(t) +R_n(t) = N_n$ is a constant upto small deviations on account of the imbalance between influxes and outfluxes. These equations extend the SIR model to a network in which population is well mixed only within each city. 
These equations provide one of the well-studied among several models of large scale infection spread \cite{TDA-spreading, Shortest-path-SIR}

Note that the Eqs.~\eqref{eq:SIR_traffic} are not India specific and have been applied on a global scale \cite{Science-Deff,SIR-meta2}.
In rest of the work, Eqs.~\eqref{eq:SIR_traffic} will be the central framework supplemented with India-specific  traffic matrix. Estimating ${\mathbf F}$ is particularly difficult due to the insufficient availability of real data, the details of which are elaborated in the supplement \cite{supplement}.

\section{Transportation network and data}
\label{chap:data}

Hereafter the discussion will be India specific. We include air, rail and road data in the traffic matrix; inland waterways and other modes are ignored. A directed network of cities/towns with a population above 1 Lakh (according to 2011 census \cite{census}) and having at least one of air, rail or road connectivity is created. The network has $M=446$ nodes (cities/towns), and 46448 weighted edges. Each pair of cities can have upto two oppositely directed edges between them, with weights representing the total traffic (all modes) in that direction. Further details of the edge data are given in the supplement \cite{supplement}.

\begin{figure}[!h]
\centering    
\includegraphics[width=0.45\columnwidth]{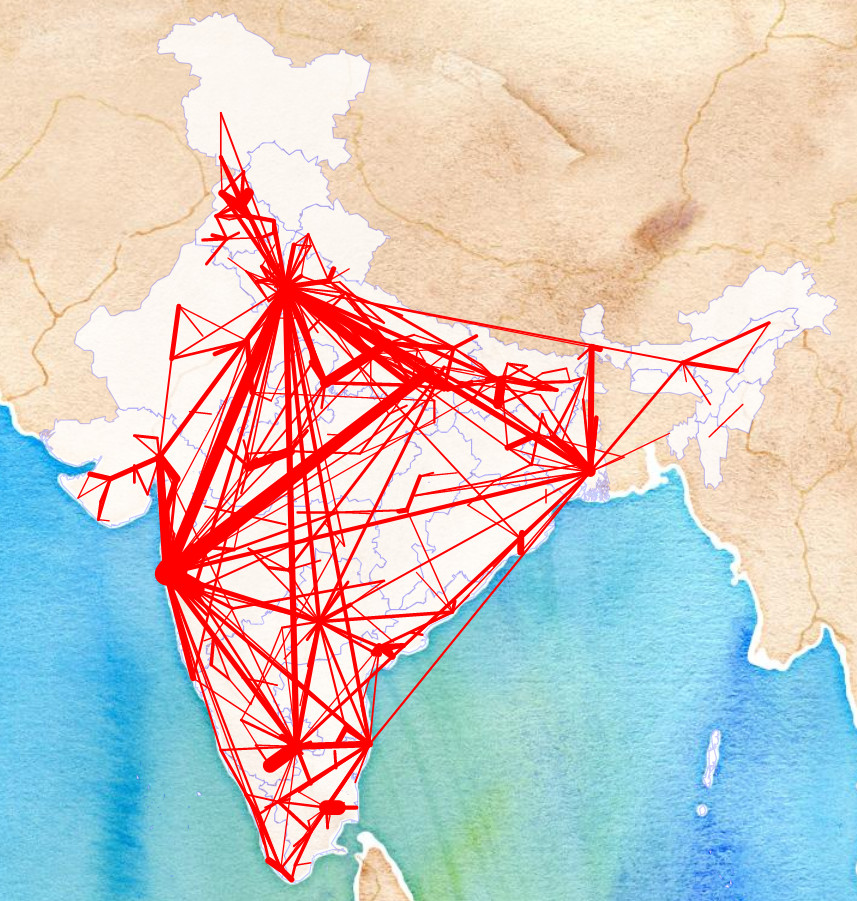}    
\caption{An averaged composite transportation network estimated based on data from pre-covid years 2017 to 2019. The lines represent the busiest 500 connections between cities and their thickness is proportional to the total volume of traffic in the forward and backward directions between each pair of cities.}
\label{fig:network_map}
\end{figure}

The air, rail and road transportation data are combined to obtain the averaged daily traffic matrix ${\mathbf F}$, whose element $F_n^m$ represents the net direct traffic (number of people) from city $n$ to $m$ on a ``typical" day. We ignore any effect of the differences in the travel times associated with different modes of traffic (for instance air travel being faster than road travel). Figure \ref{fig:network_map} shows the busiest 500 inter-city routes based on sum of forward and backward traffic. 

The matrix \textbf{F} constructed from real data is not symmetric $F_n^m\neq F_m^n$, {\it i.e.}, the forward and backward traffic between $n$ and $m$ is unequal. The line thickness indicates its weight -- thicker lines represent more traffic. 

\begin{table}[!h]
\centering
\renewcommand{\arraystretch}{1.5}
\begin{tabular}{|l|c|c|c|c|}
\hline
Property & \multicolumn{1}{c|}{Air} & \multicolumn{1}{c|}{Rail} & \multicolumn{1}{c|}{Road} & \multicolumn{1}{c|}{Combined} \\ [2.0mm]\hline 
Number of nodes & 85 & 435 & 446 & 446 \\ [2.0mm]\hline
Number of edges & 1182 & 41594 & 9128 & 46448 \\ [2.0mm]\hline
Average degree & 13 & 95 & 20 & 104 \\ [2.0mm]\hline
Passengers per day & $7.5 \times 10^5$ & $8.8 \times 10^6$ & $2.5 \times 10^6$ & $1.2 \times 10^7$ \\ [2.0mm]\hline
Fraction of total & 0.06 & 0.73 & 0.21 & 1.0 \\ [2.0mm]\hline
\end{tabular}
\caption{Properties of the transportation network and mobility data assembled for this work. Not surprisingly,
air travel constitutes a small fraction of the overall mobility. Majority of the long-distance travel is accounted for by trains and road is the dominant mode at short distances.}
\label{table:trafficdata}
\end{table}

\section{Infectious diseases hazard index}
\label{inf_dis_haz_map}
The central idea in constructing the hazard map is the notion of \textit{effective distance}
introduced in Ref. \cite{Science-Deff}. If $F^n$ and $F_n$, are the net rates of people traveling in and out of city $n$, then the one-step conditional probability that a person leaving city $n$ travels to $m$ is given by,
\begin{align}
P_n^m = \frac{F_n^m}{F_n}.
\end{align}
We define pair-distance $d_n^m$ from city $n$ to $m$ as,
\begin{align}
d_n^m = 1- \log{P_n^m}.
\end{align}
If the cities are not directly connected {\emph i.e.} nobody travels from $n$ to $m$ directly, 
then $P_n^m =0$ and $d_n^m \rightarrow \infty$. In contrast, large traffic between the cities (relative to the population of the origin $n$) makes $d_n^m$ small. Note that $d_n^m$ is not necessarily symmetric between the cities. 

Fastest path for an infection may pass through other cities. This motivates the notion of an effective shortest distance between any pair of cities as follows.
For any path $\Gamma_n^m$ (a sequence of cities starting and ending at $n$ and $m$) through the network between $n$ and $m$, $\lambda(\Gamma_n^m)$ represents the sum of the pair distances between successive cities.
The effective distance $D_{\rm eff}^{nm}$ between a pair of cities is defined as the shortest among all  paths $\Gamma_n^m$:
\begin{align}
D_{\rm eff}^{nm} = \min_{\text{ all paths }\Gamma_n^m} \; \lambda(\Gamma_n^m)\label{eq:Deff}.
\end{align}

Infection spread between the cities is likely to depend on the traffic between cities rather than geographical distances. 
The effective distance depends on the mobilities rather than the geographical distance. By definition it takes into account the multiple paths that may connect a pair of cities, just as the infection may reach a city through another one rather than directly from the outbreak. It is therefore natural to expect a high correlation between aspects of infection spread and $D_{\rm eff}$.

To make this precise, we define the ``time of arrival" $T_A^{nm}$ of the infection in a city $m$ (from a given outbreak location $n$) as the first time when the number of (active) infected cases cross a predefined threshold $I^c$. 
In studies of infection spread through global air-traffic patterns, time of arrival at location $m$ from an outbreak location $n$ was found to be proportional to the effective distance $D_{\rm eff}^{nm}$ between them.\cite{Science-Deff}.
Naively $T_A^{nm}$ between cities, which is obtained from solution of the Eq.\ref{eq:SIR_traffic} is expected to have a complex dependence on the traffic. It is surprising that $T_A^{nm}$ can instead be reliably predicted from a simple functional $D_{\rm eff}^{nm}$.

Extensive simulations performed in this work and summarized in Fig.~\ref{fig:deff_vs_dgeo} show that, for a wide range of realistic $\alpha$ and $\beta$ and Indian traffic patterns, $T_A$ has a high linear correlation with $D_{\rm eff}$.
Predicting the arrival of infection at a given location is not only of academic interest but is also of immense practical value. In the rest of the paper, we present our analysis of this idea using Indian traffic data.
%

Given an outbreak location, the risk of infection in another location can be quantified in many ways, time of arrival being a natural one. Under reasonable and realistic assumptions including uniform infection parameters, $D_{\rm eff}$ provides a reliable and robust predictor of $T_A$, as evident from our simulations. $D_{\rm eff}$ can be mapped using available transportation data unlike $T_A$ that can be obtained either from extensive simulations or a posteriori knowledge of the infection spread.

\section{Results}
\label{chap:results}

\begin{figure}[!h]
    \centering    
    \includegraphics[width=0.38\columnwidth]{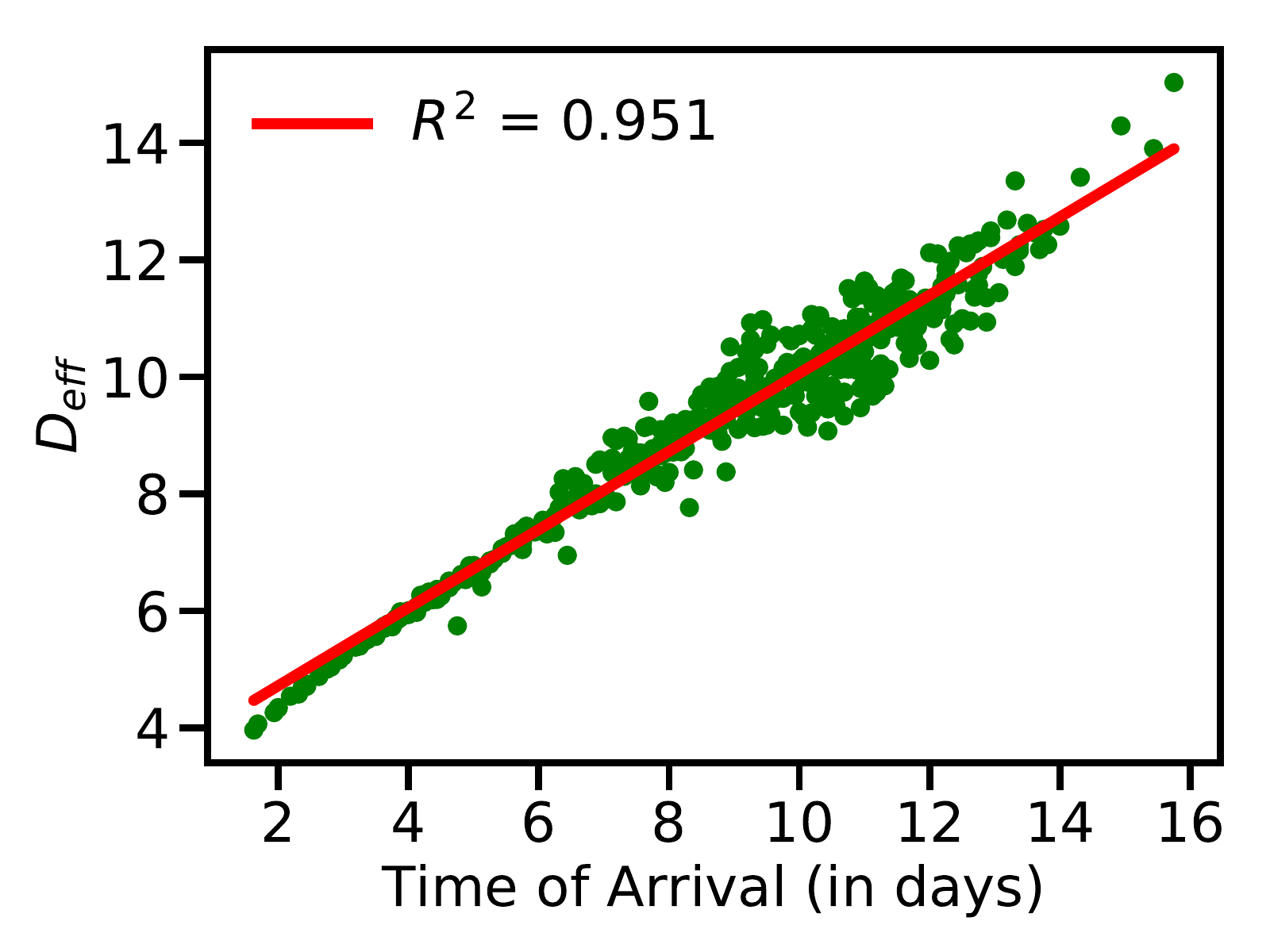}\includegraphics[width=0.38\columnwidth]{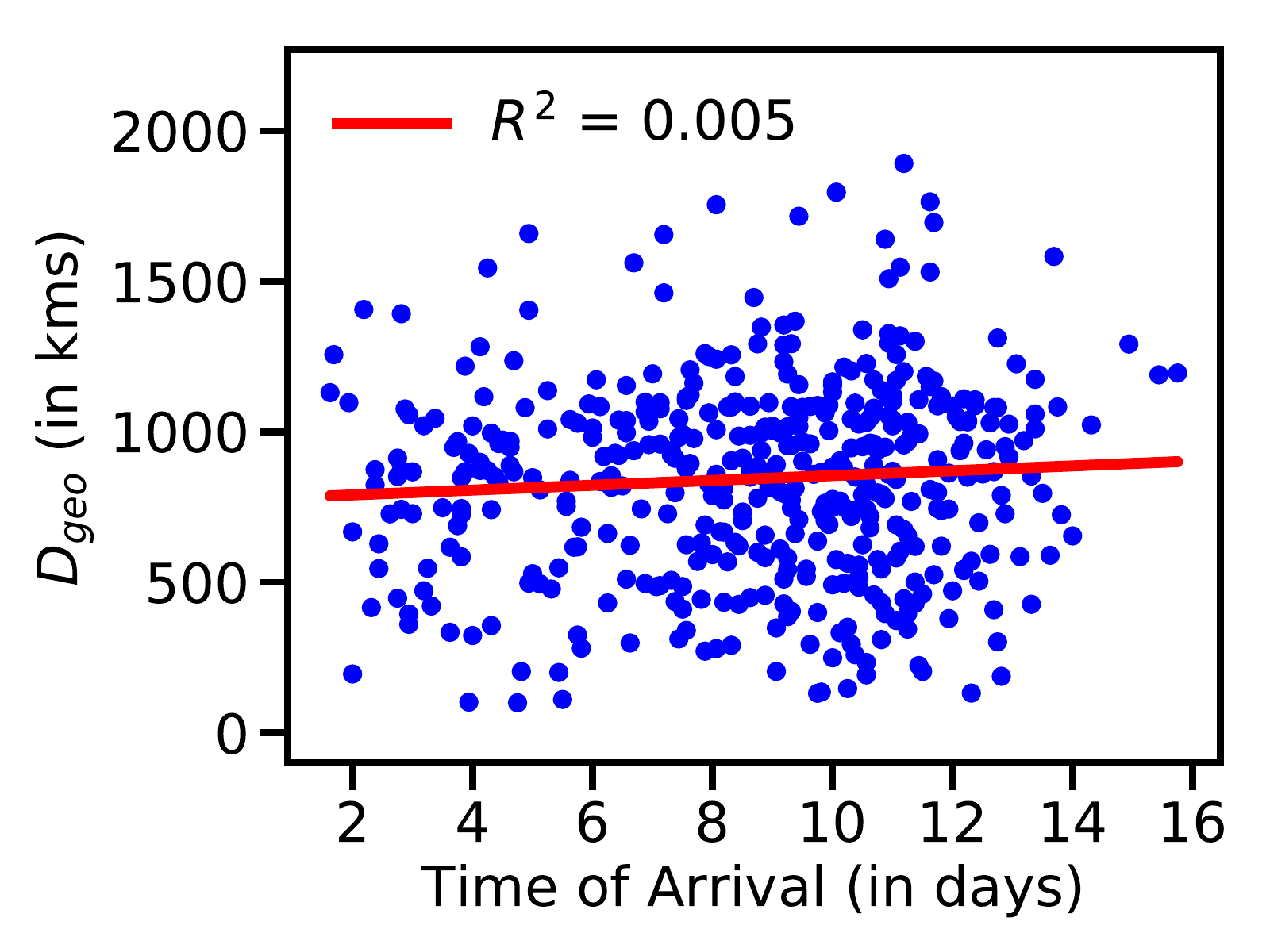}
    \includegraphics[width=0.38\columnwidth]{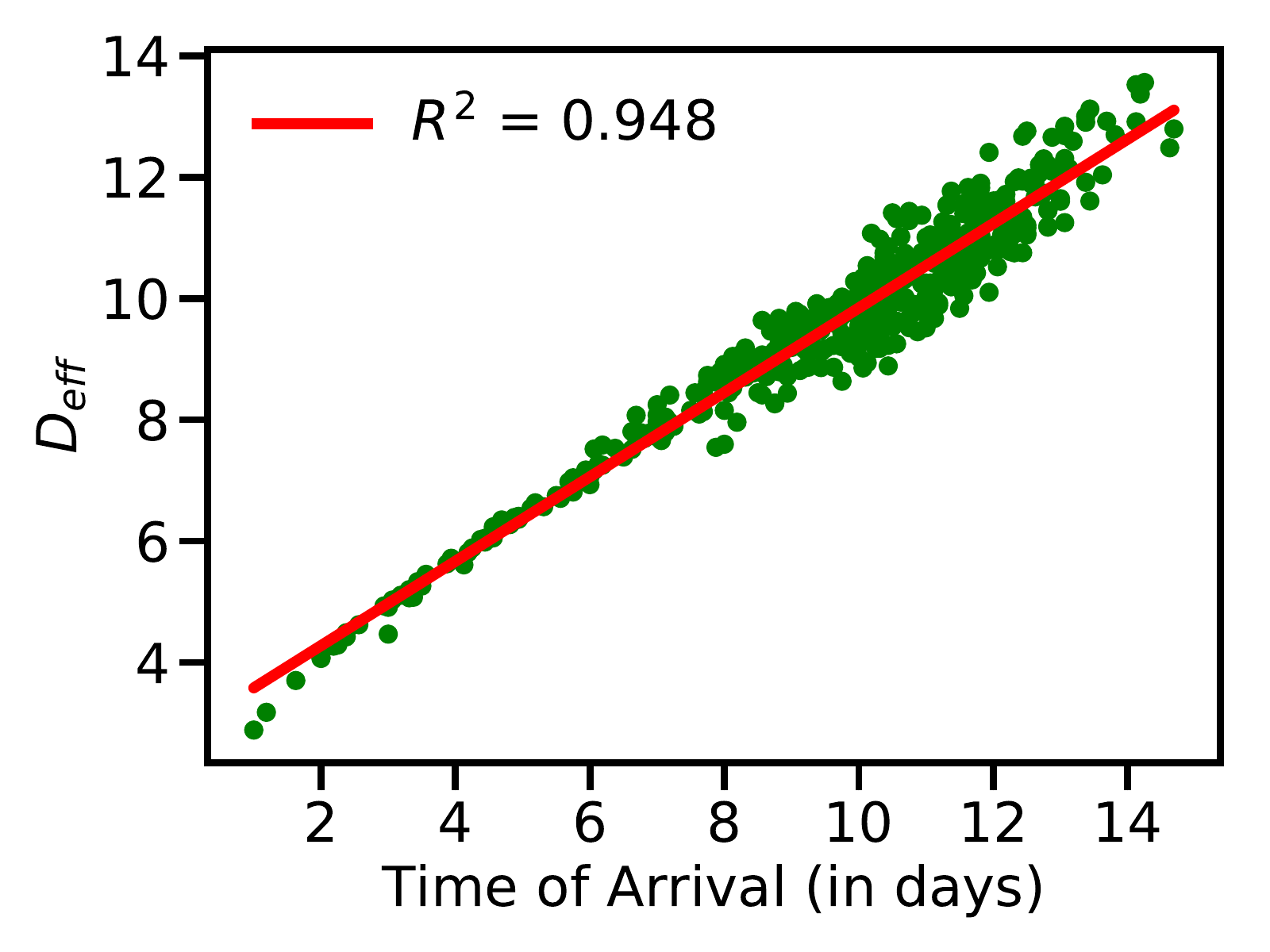}\includegraphics[width=0.38\columnwidth]{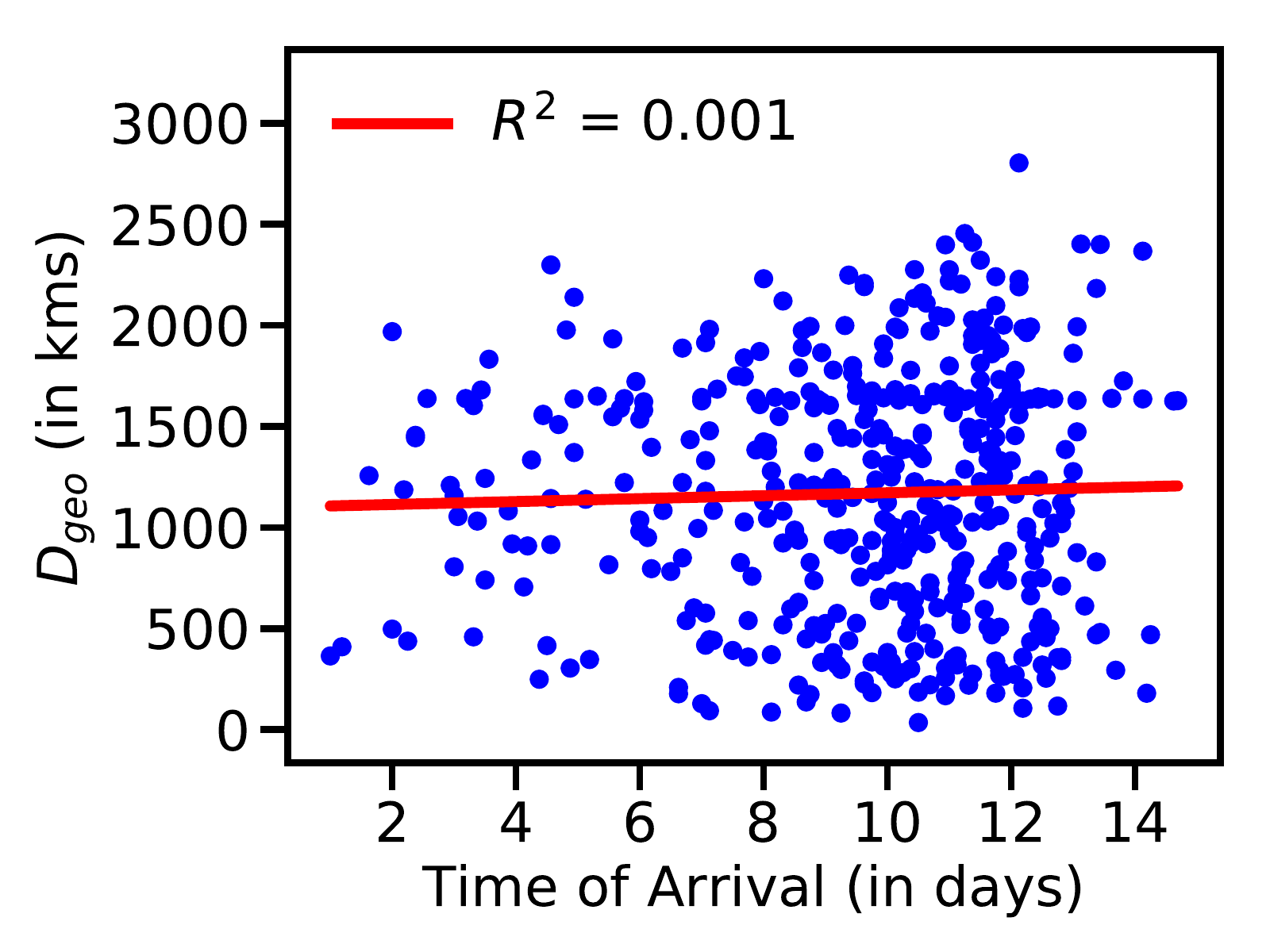}
    \includegraphics[width=0.38\columnwidth]{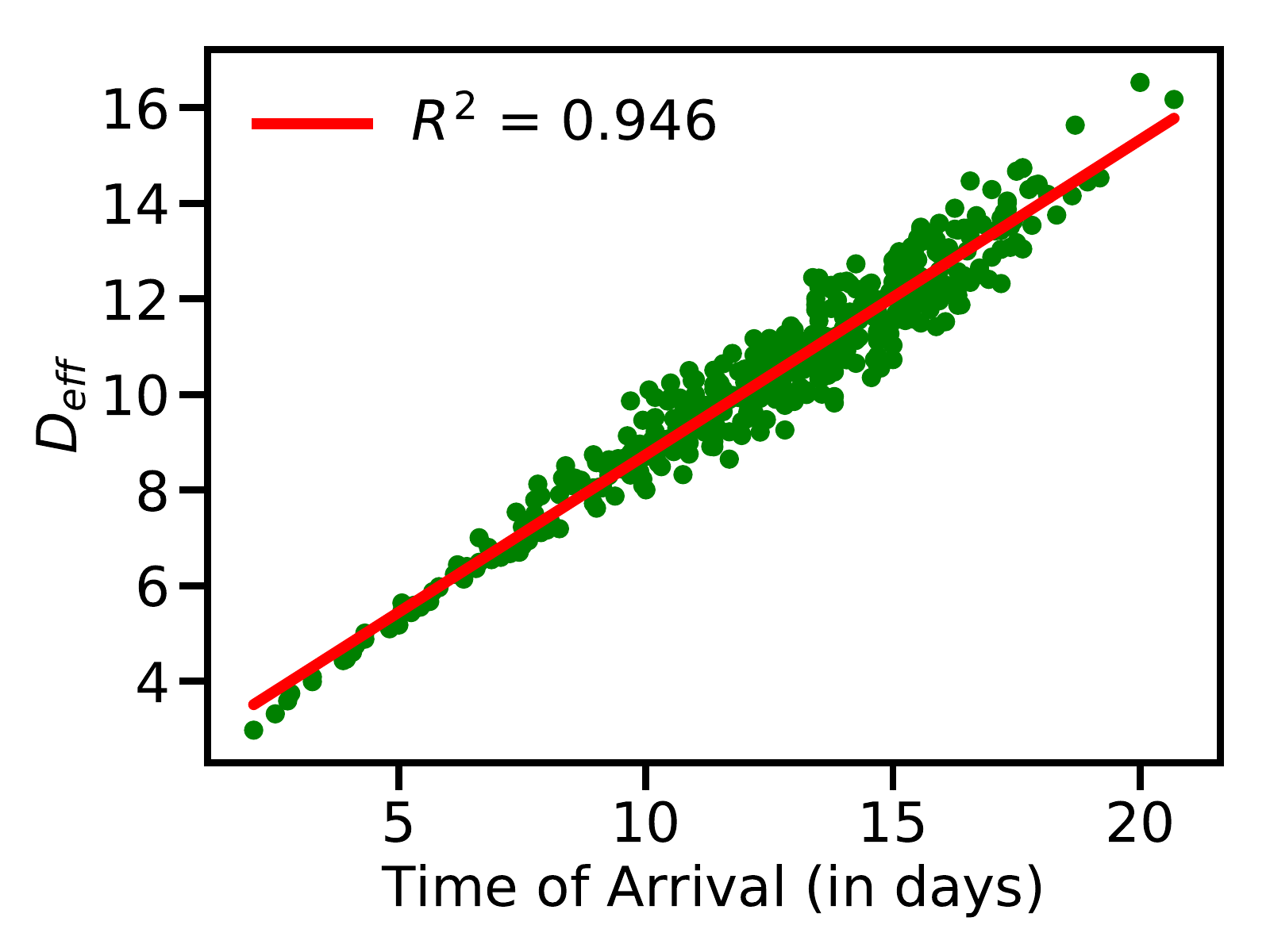}\includegraphics[width=0.38\columnwidth]{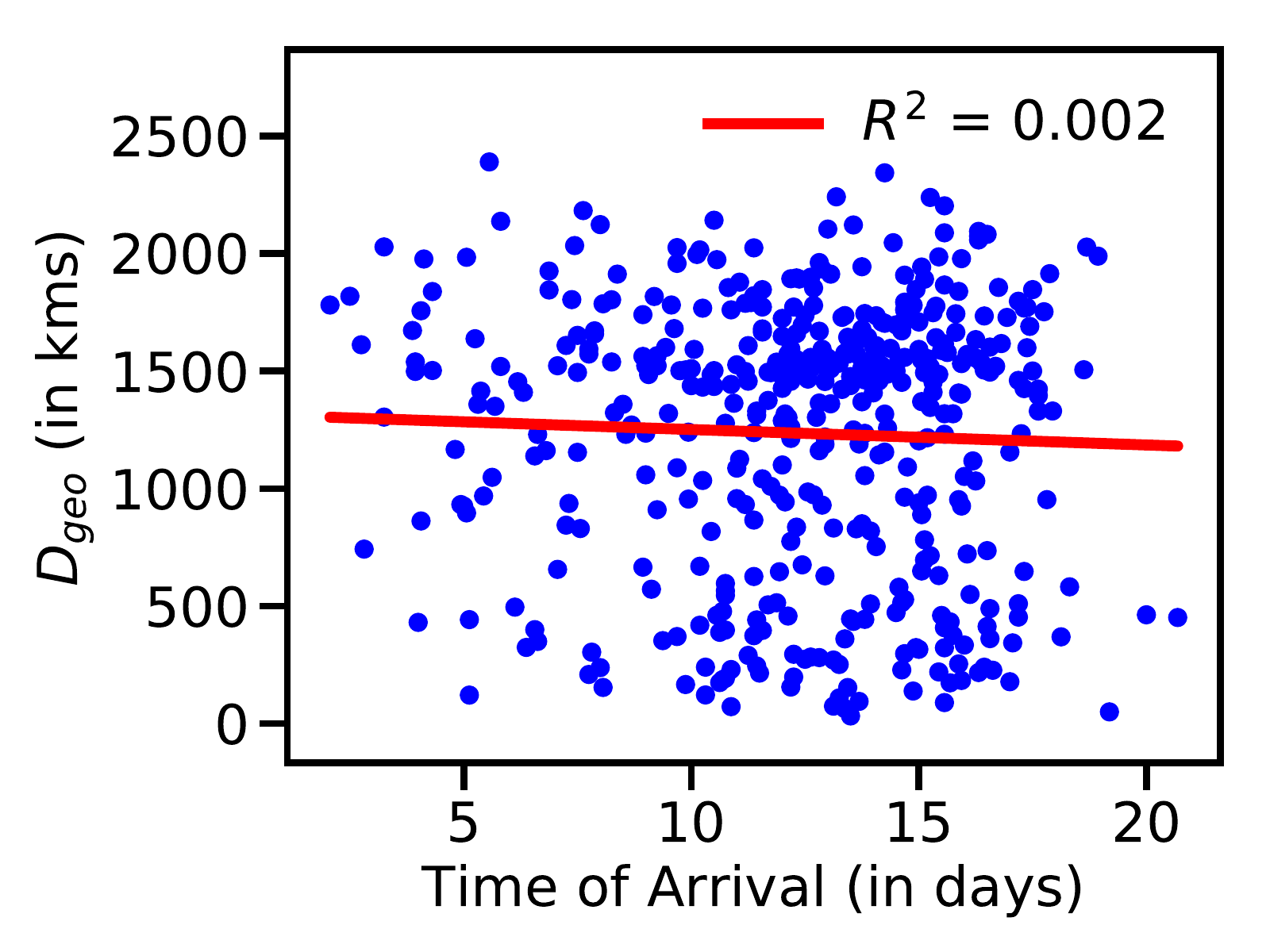}
    \includegraphics[width=0.38\columnwidth]{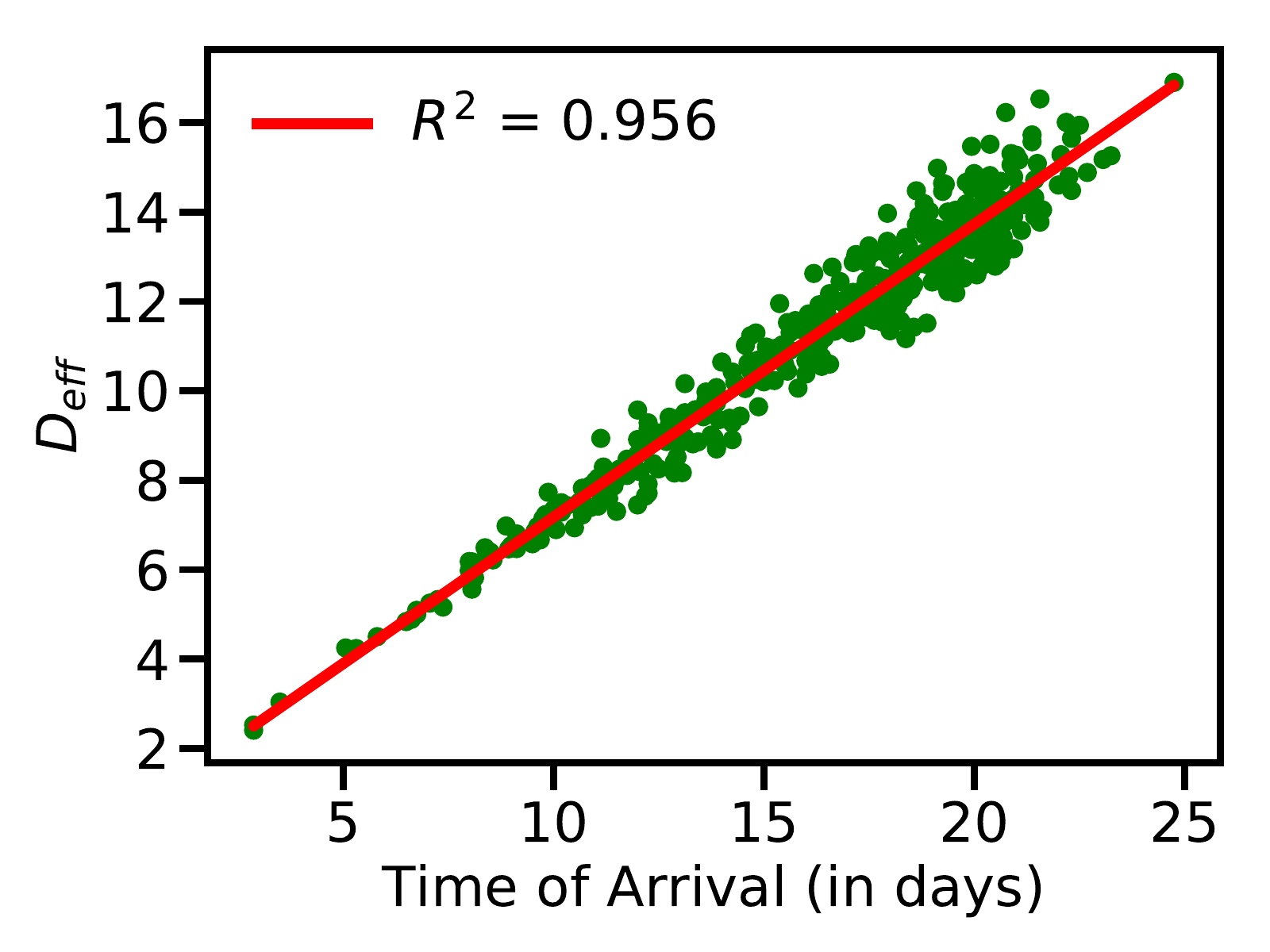}\includegraphics[width=0.38\columnwidth]{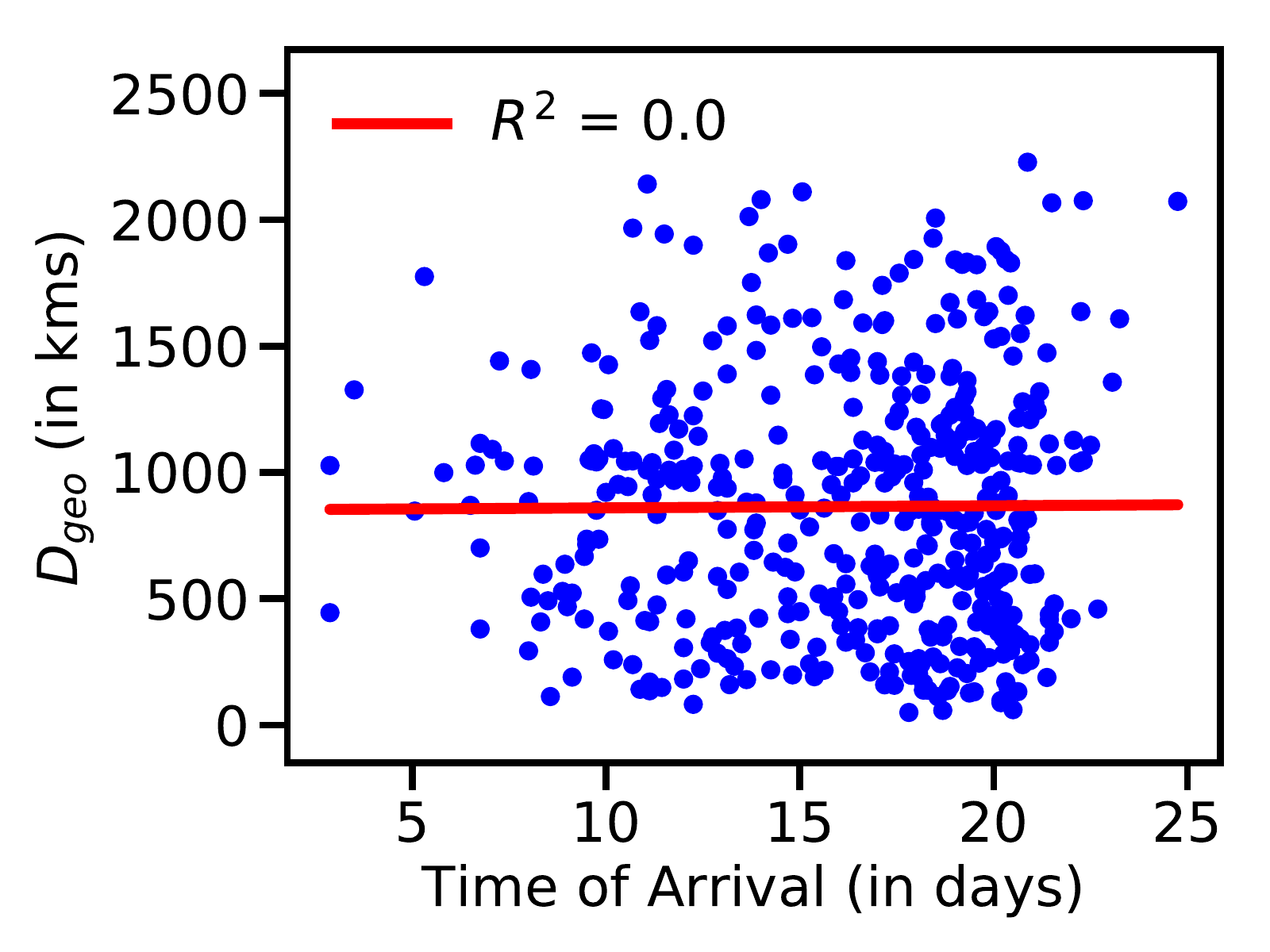}
    
\caption{Plots showing strong linear correlation between effective distance $D_{\rm eff}$ and time of arrival $T_A$.
Left column shows $D_{\rm eff}^{i_0 m}$ plotted against $T_A$ of the infection at city $m$ from outbreak at city $i_0$. 
$T_A$ is obtained by solving Eq.~\eqref{eq:SIR_traffic} with infection parameters $\alpha=1.5, \beta=1.0, I^c=10$. 
Outbreak locations considered in the rows from top to bottom are Delhi, Mumbai, Patna, and Tirupati. 
The right column shows the geographical distance $D_{\rm geo}^{i_0 m}$ from $i_0$ plotted against $T_A$.
$R^2$ indicated in the plots is a measure of goodness of the linear fits (red lines), $R^2=1$ being a perfect linear fit.
}
    \label{fig:deff_vs_dgeo}
\end{figure}

In order to validate the utility of the effective distance, we numerically evolve the coupled differential equations \eqref{eq:SIR_traffic} using fourth-order Runge-Kutta method. The initial  infected population $I_{i_0}(t=0)$ in the outbreak city $i_0$ is taken to be a fraction ($0.0001$) of the local population. 
We perform such simulations for different choices of the outbreak locations and infection parameters. $T_A$ for each city is evaluated in each case by finding the time when the city's infected population crosses a threshold, $I^c$ taken to be $10$. Qualitative results are independent of choices of $I_{i_0}$ and $I^c$.

In Fig.~\eqref{fig:deff_vs_dgeo} we show the results assuming infection parameters $\alpha=1.5, \; \beta=1.0$  giving $R_0=1.5$, a typical value that was witnessed for SARS-CoV-2 \cite{r0info}.
In Fig.~\eqref{fig:deff_vs_dgeo} (left panel), the effective distance $D^{i_0m}_{\rm eff}$, where $i_0$ is the outbreak location, is plotted against the time of arrival at city $m$.
This is shown for four different outbreak locations of varying sizes,
namely, Delhi, Mumbai, Patna, and Tirupati. 
We find a good linear relation between $D^{i_0m}_{\rm eff}$ and $T_A^{i_0m}$, as indicated by high $R^2 \gtrsim  0.94$. These are in striking contrast to the right panels which show $T_A$ against the geographical distances from the outbreak. 

Similar observations were made in Ref.\cite{Science-Deff} which considered key global air traffic patterns alone. Remarkably, within India, considering multiple modes of transport, with air travel being the least popular mode accounting for less than 10\% of relevant mobility, the linearity holds good. Smaller $D_{\rm eff}$ to the outbreak then suggests a higher risk to a city, manifested as earlier arrival of the infection. The demonstration of the ability of the $D_{\rm eff}$ to predict the time of arrival is a key result of this work.

\begin{figure}[t]
    \centering    
    \includegraphics[width=0.45\columnwidth]{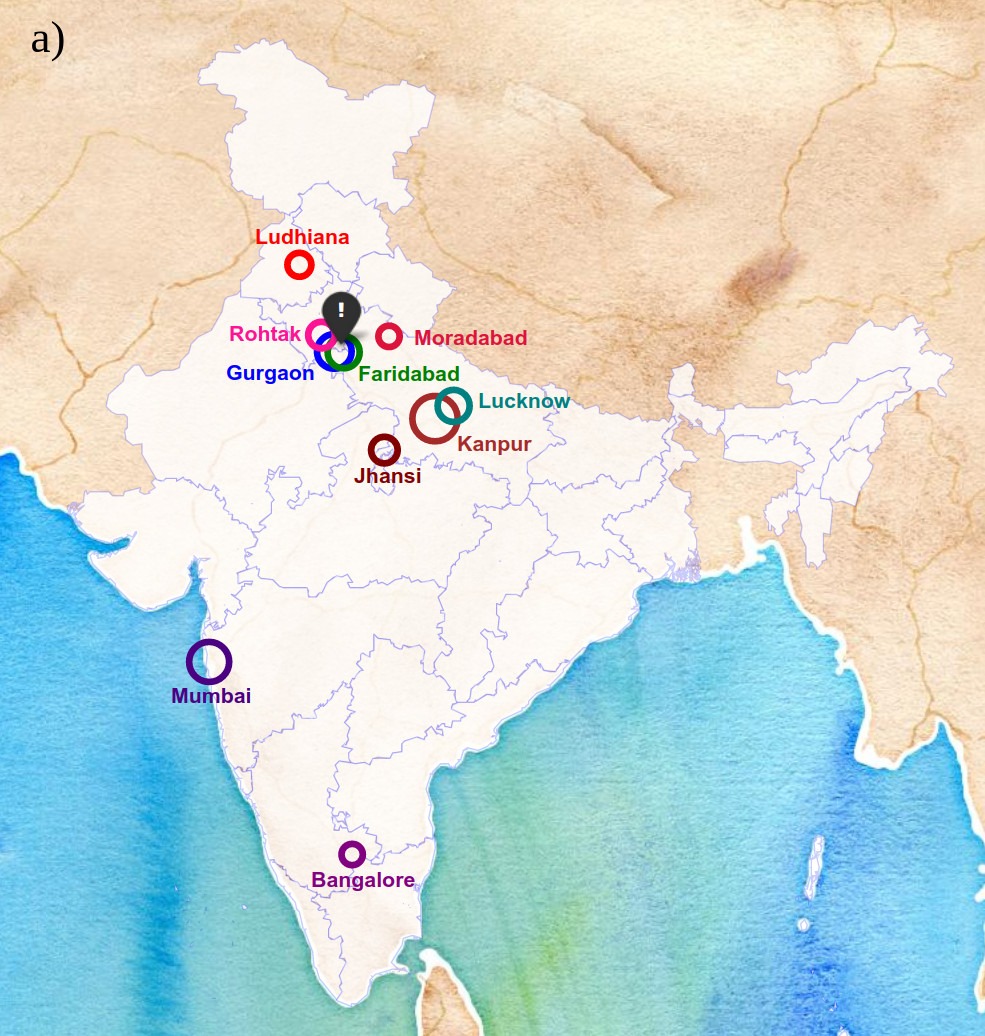}\hspace{1.5em}\includegraphics[width=0.45\columnwidth]{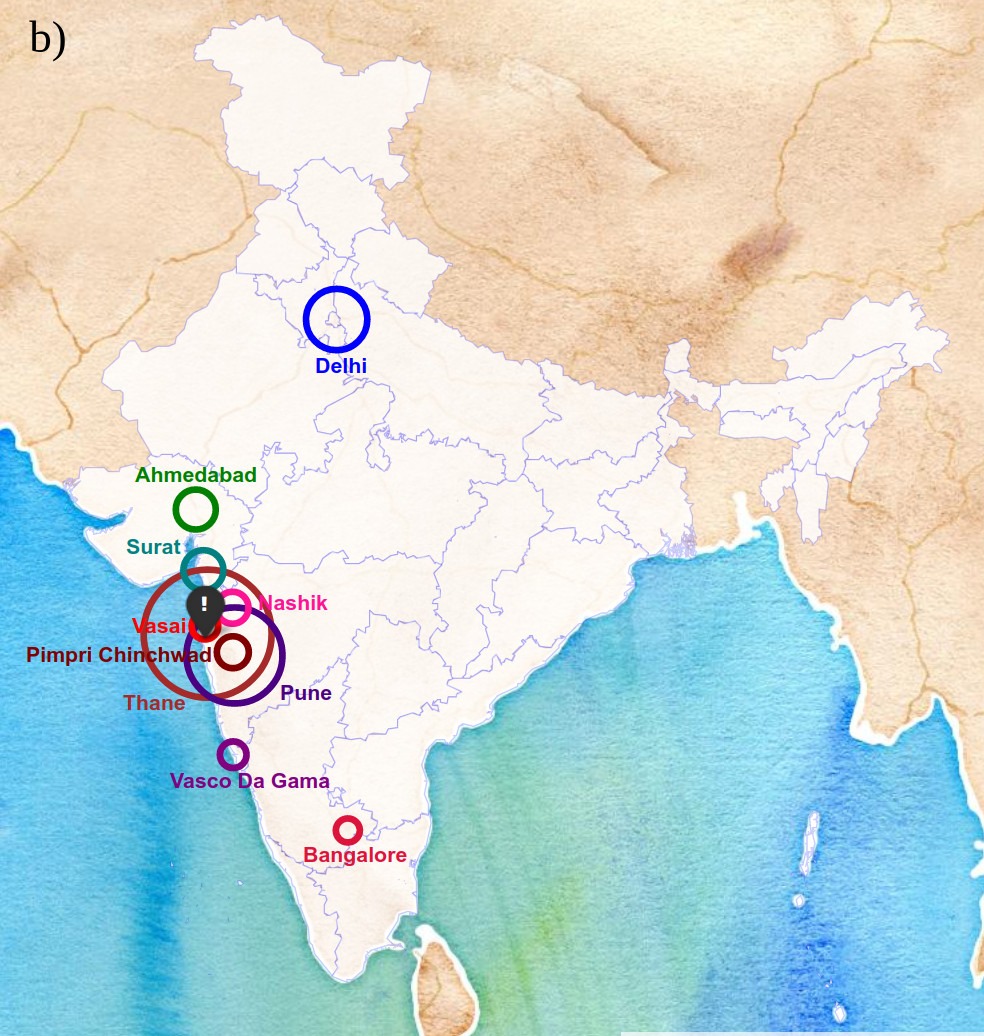}\vspace{1.5em}
    \includegraphics[width=0.45\columnwidth]{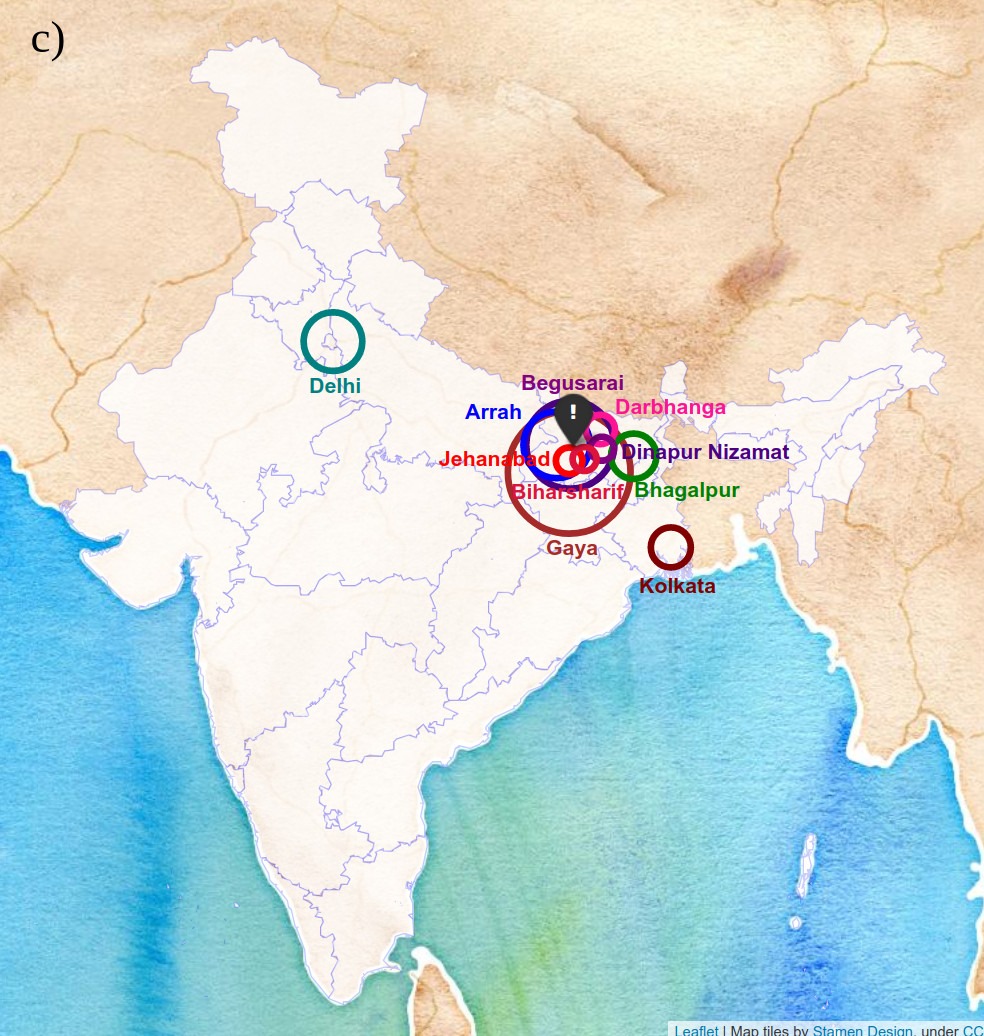}\hspace{1.5em}\includegraphics[width=0.45\columnwidth]{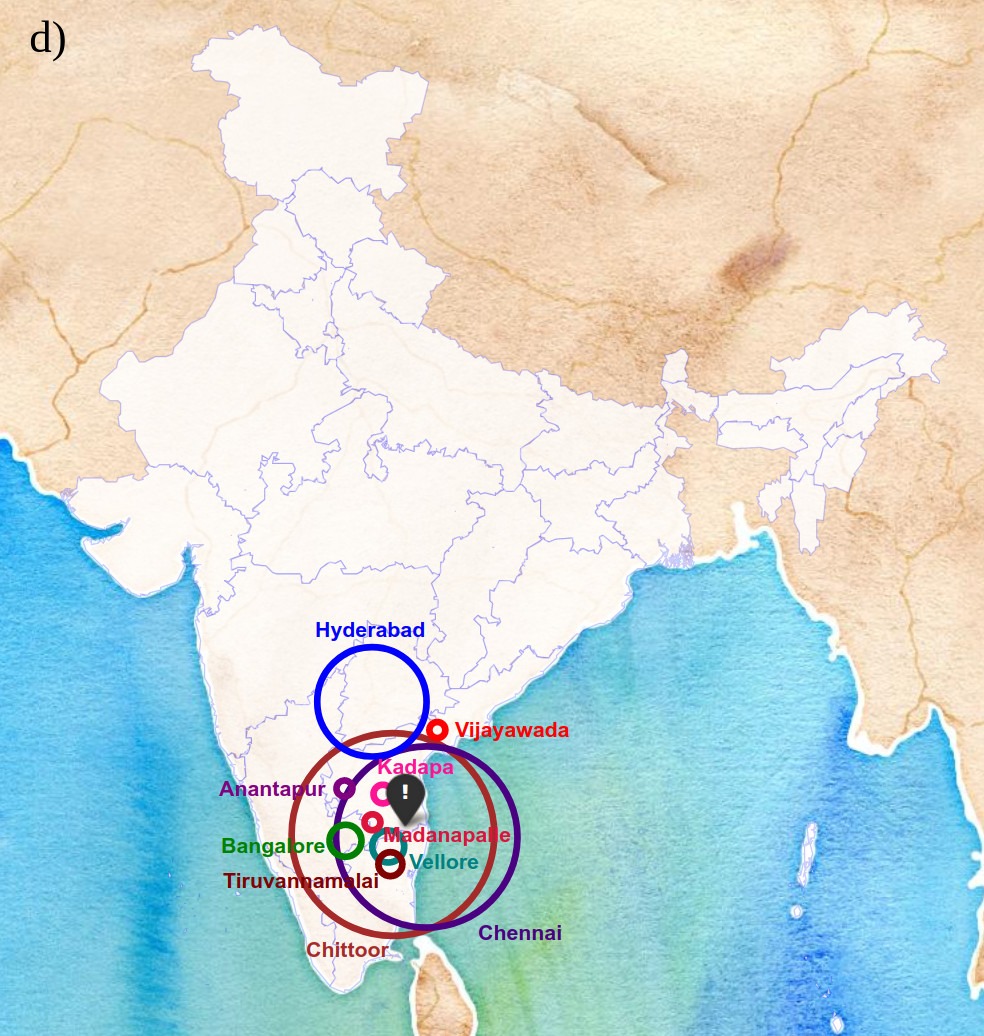}
     \caption{A visual depiction of the information in Table \eqref{table:4cities_hi} in the form of an infectious diseases hazard map, with outbreak locations at a) Delhi, b) Mumbai, c) Patna and d) Tirupati (shown as black colored location icon). The radius of circle is proportional to the hazard index of the city/town. Larger the circle, greater is the hazard and their color does not carry any information. Only the cities/towns with top-10 hazard values are shown.}
    \label{fig:hmap1}
\end{figure}

Table \ref{table:4cities_hi} shows the $T_A$ for the same outbreak locations as in Fig. \ref{fig:deff_vs_dgeo}. 
In each case, the list of the top cities in terms of risk ({\emph i.e.} smallest $D_{\rm eff}$) is also listed.
For outbreaks from poorly connected cities, the surrounding regions face the first brunt of infection; followed by bigger cities. 
On the other hand, outbreaks from big metros which are well connected, quickly reach far corners. For instance, infection from Tirupati reaches Bangalore in $\sim 5$ days, while outbreak from Mumbai or Delhi, spreads to Bangalore in $\sim 2.5$ days.
The hazard map in Fig. \ref{fig:hmap1} shows this visually for the same four outbreak locations. The size of circles represent the hazard (larger the circle, greater the risk). 
The hazard ({\emph i.e} $D_{\rm eff}$) is easily estimated for all the cities; only the top 10 cities are shown to avoid clutter.

\begin{figure}[t]
\centering    
   \includegraphics[width=0.45\columnwidth]{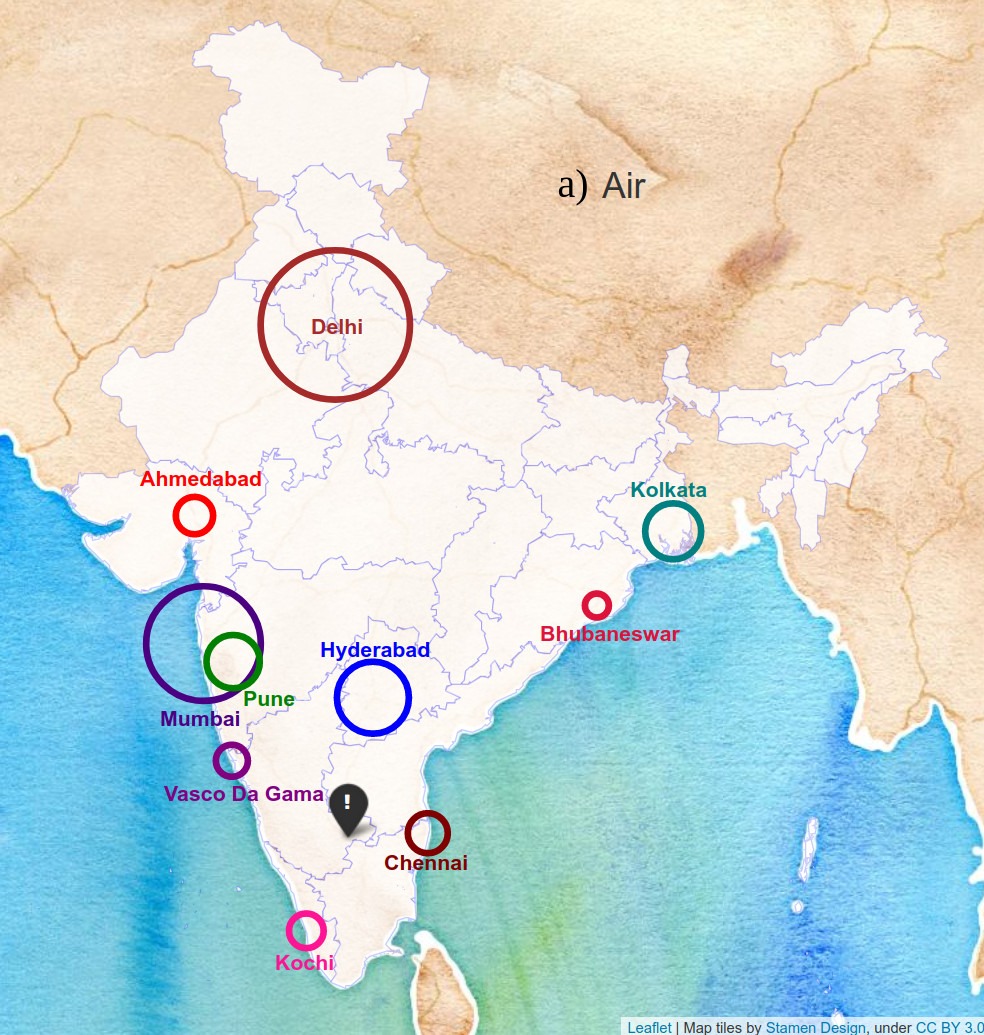}\hspace{1.5em}\includegraphics[width=0.45\columnwidth]{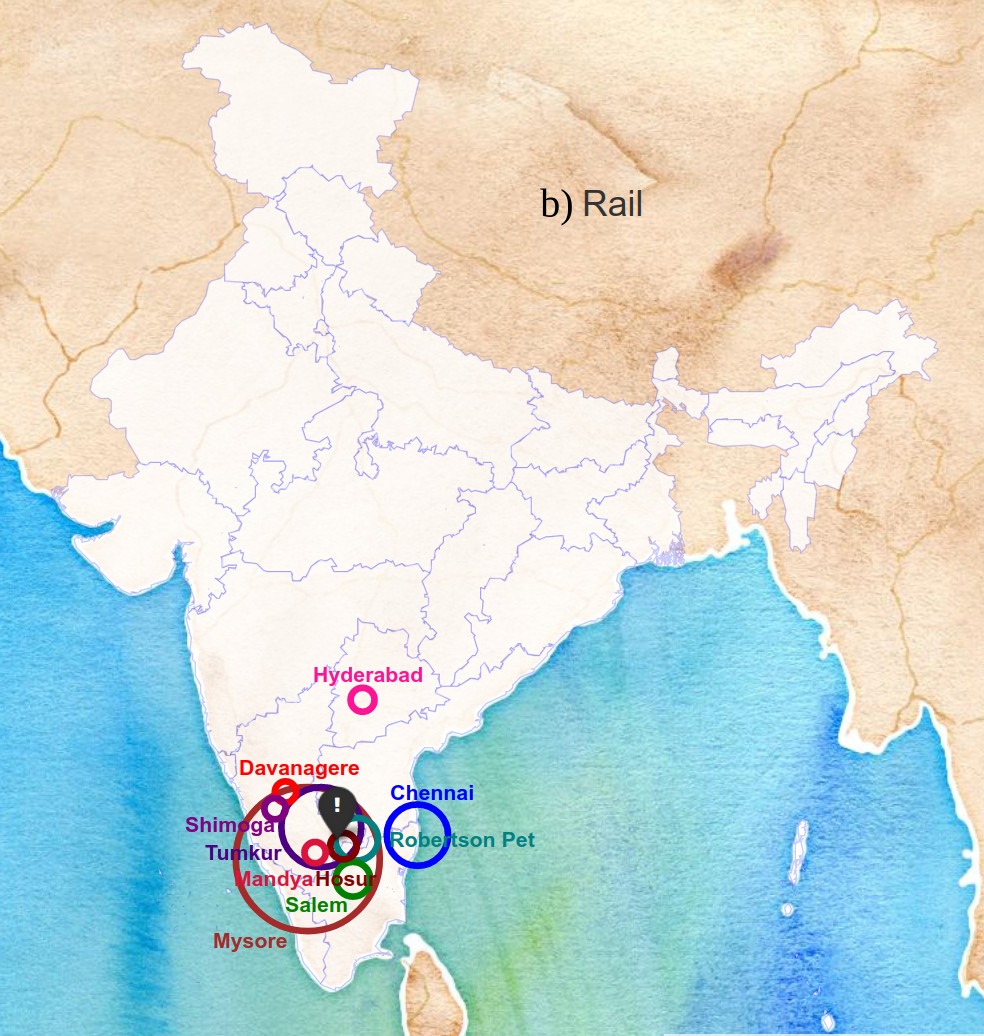}\vspace{1.5em}
    \includegraphics[width=0.45\columnwidth]{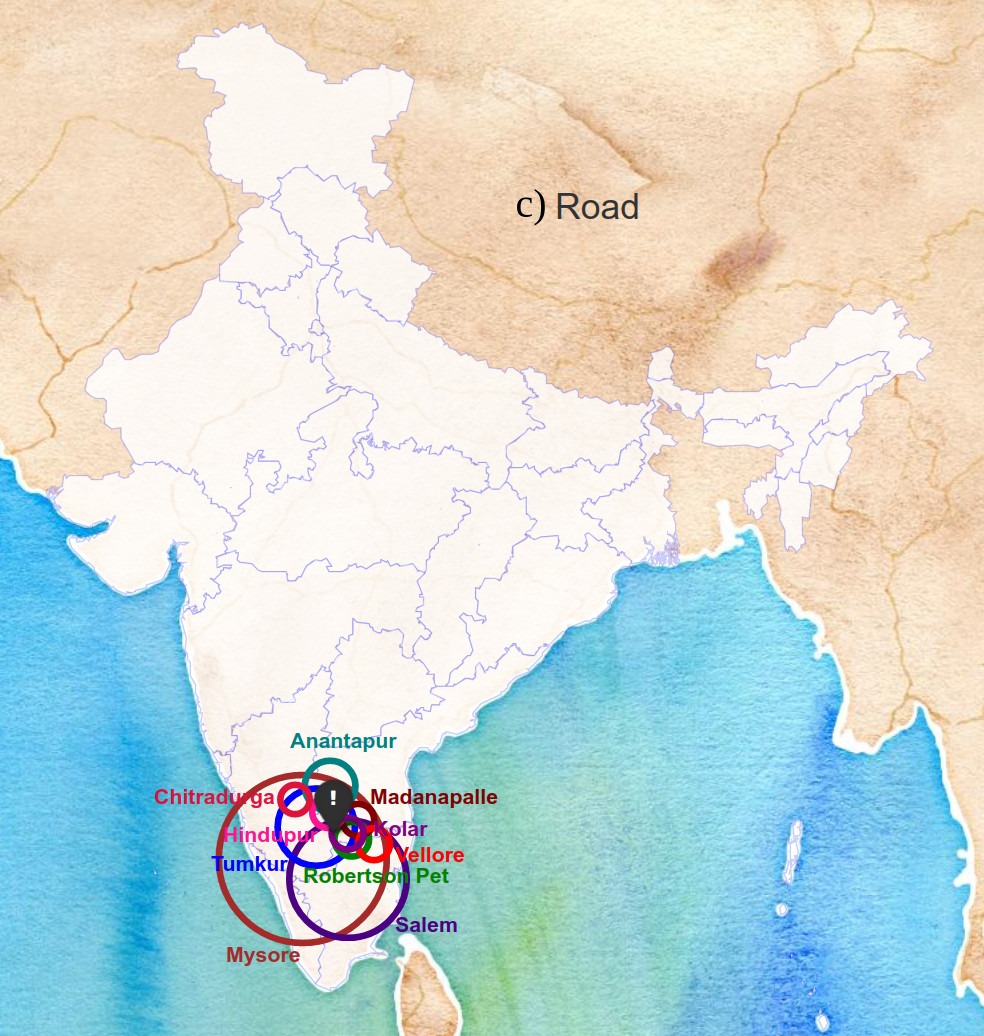}\hspace{1.5em}\includegraphics[width=0.45\columnwidth]{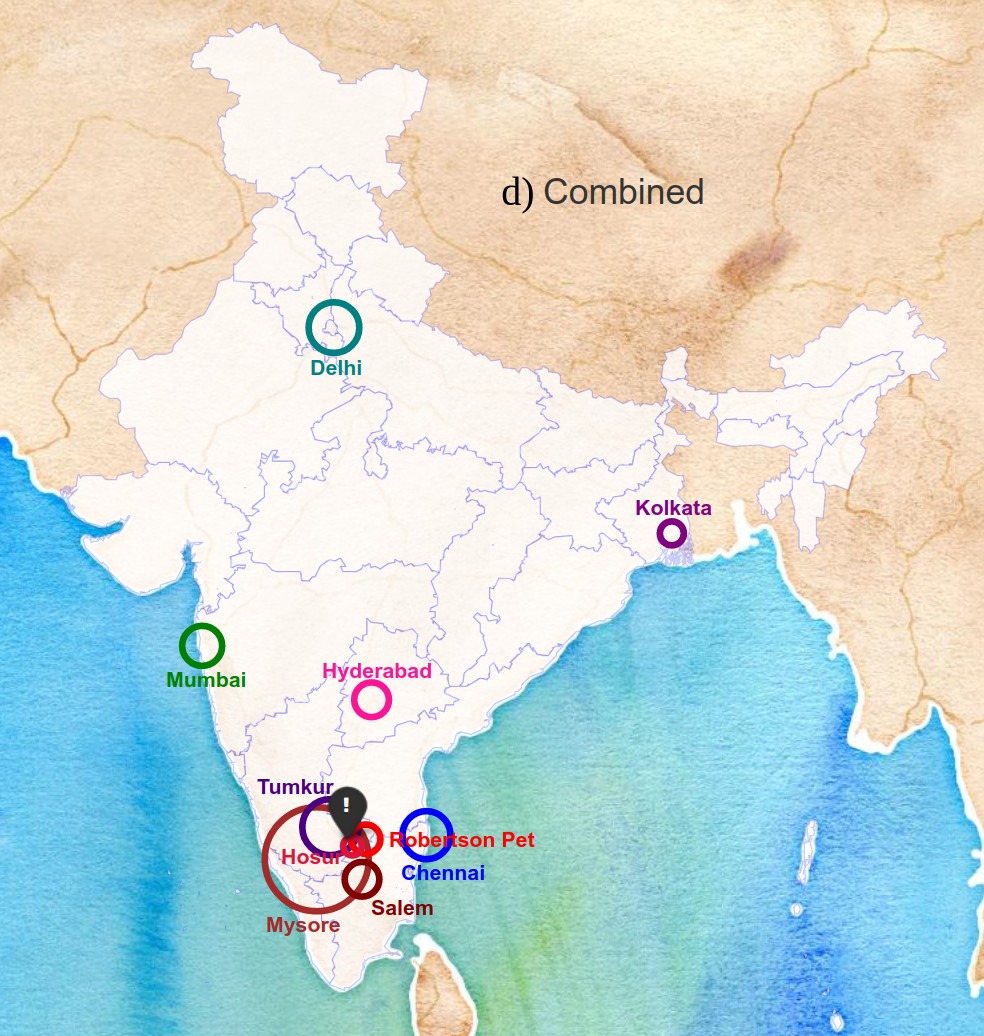}
     \caption{Transportation mode specific Hazard maps with Bangalore as outbreak location. The figures correspond to a) Air, b) Rail,  c) Road, and d) Combined modes of transport. The radius of circle is proportional to the hazard index of the city/town. Larger the circle, greater is the hazard and their color does not carry any information. Only the cities/towns with top-10 hazard values are shown.}
    \label{fig:hmap_bangalore}
\end{figure}

\begin{figure}[t]
    \centering    
   \includegraphics[width=0.45\columnwidth]{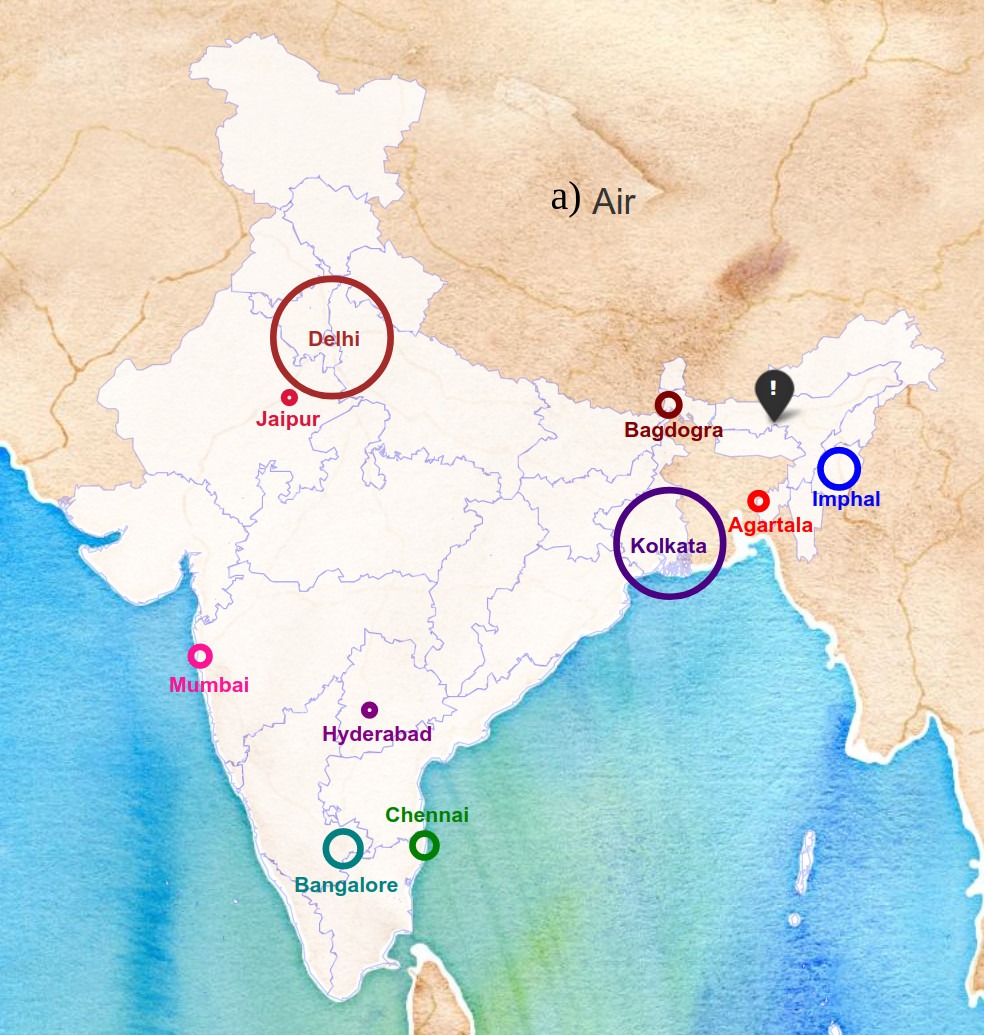}\hspace{1.5em}\includegraphics[width=0.45\columnwidth]{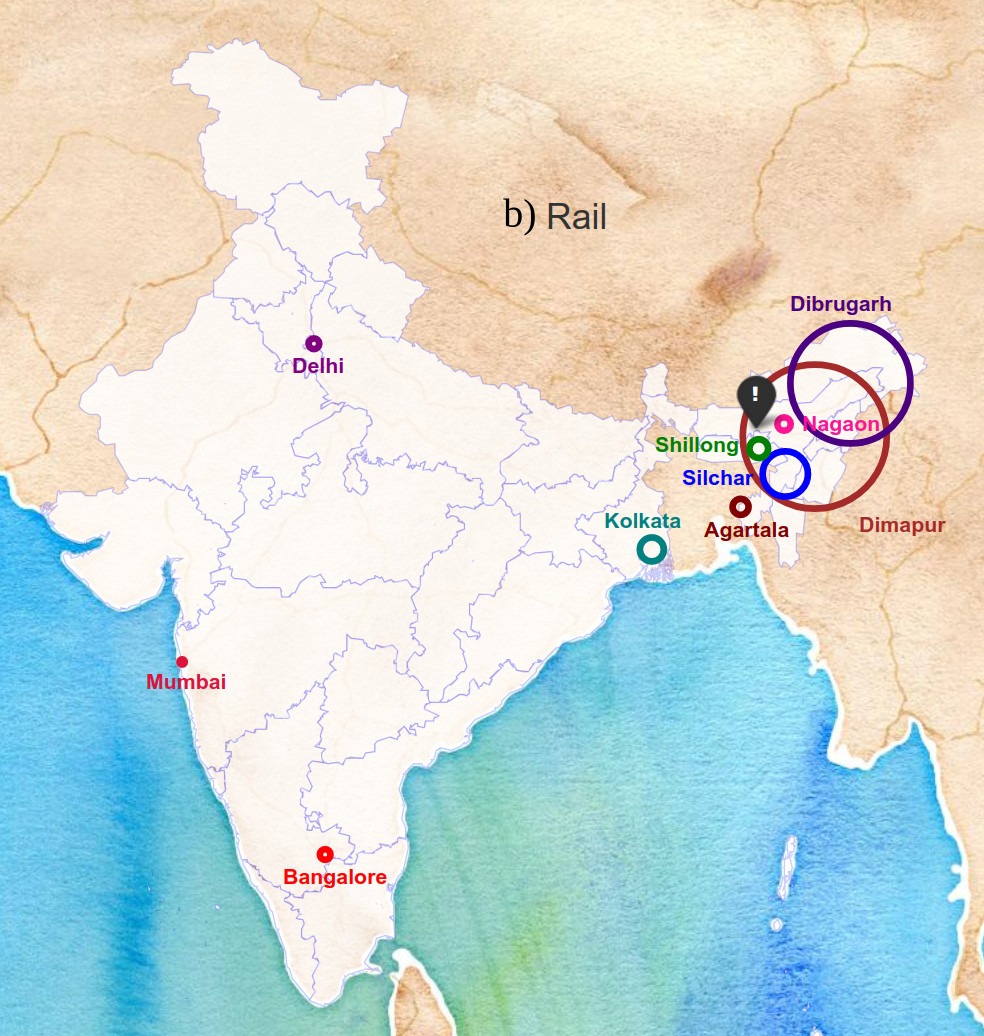}\vspace{1.5em}
    \includegraphics[width=0.45\columnwidth]{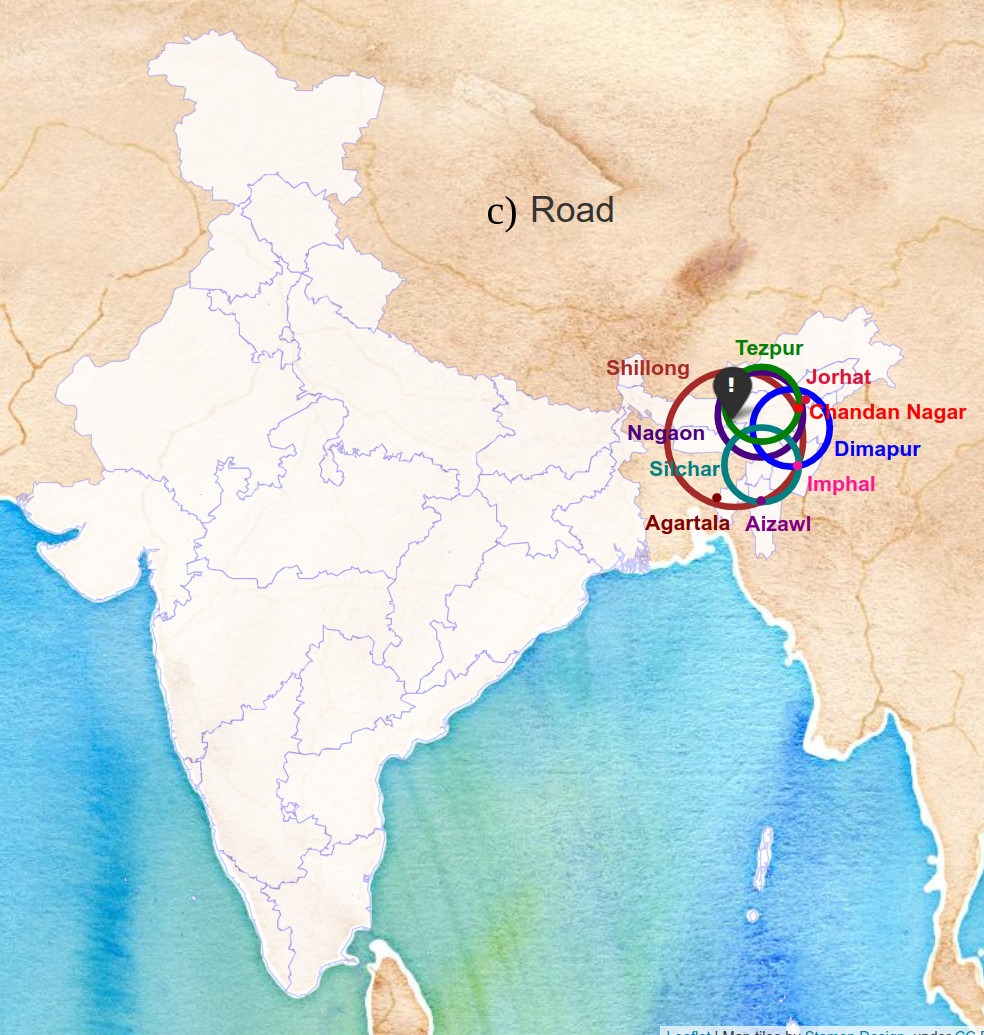}\hspace{1.5em}\includegraphics[width=0.45\columnwidth]{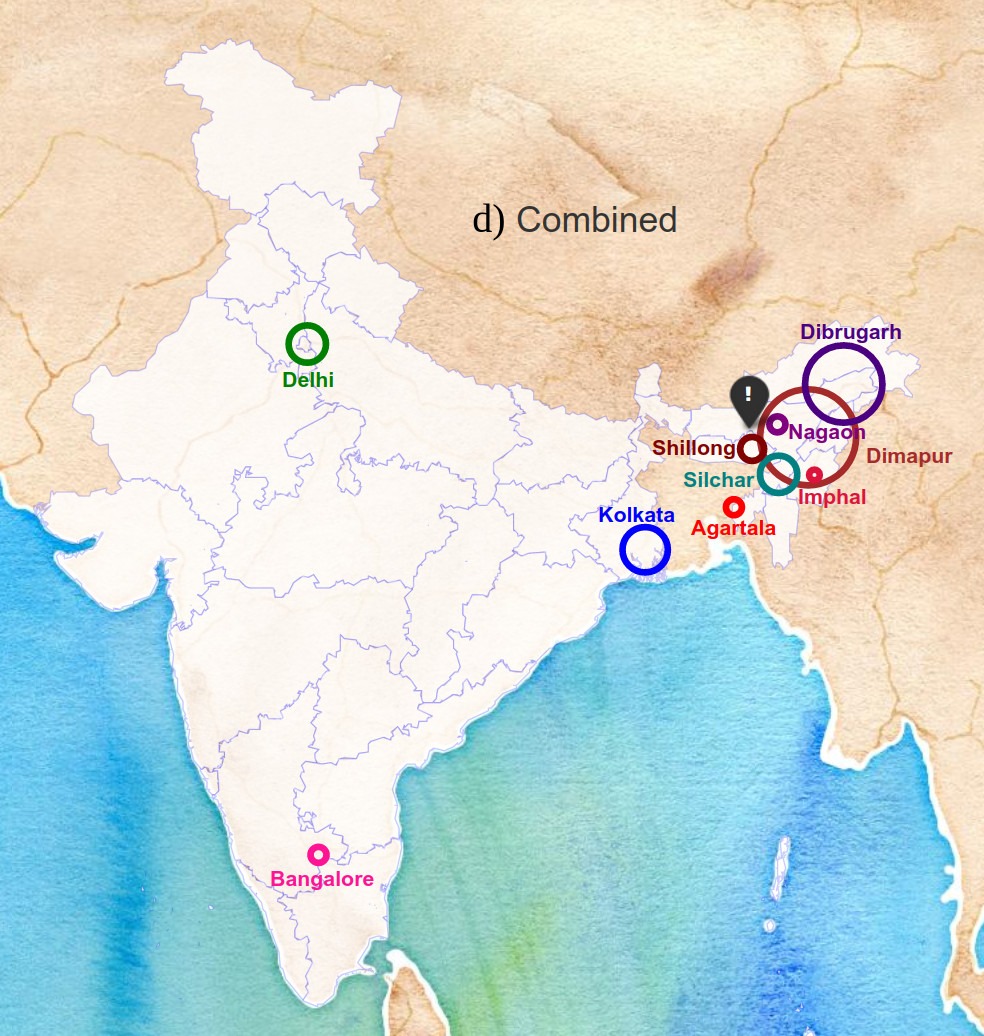}
    \caption{Transportation mode specific Hazard maps with Guwahati as outbreak location. The figures correspond to a) Air, b) Rail,  c) Road, and d) Combined modes of transport. The radius of circle is proportional to the hazard index of the city/town. Larger the circle, greater is the hazard and their color does not carry any information. Only the cities/towns with top-10 hazard values are shown.}
    \label{fig:hmap_guwahati}
\end{figure}

It is interesting to consider the hazard map assuming only one mode of transport is operating. The transportation-mode-specific hazard map is shown for two outbreak locations, Bangalore (Fig.~\ref{fig:hmap_bangalore}),and Guwahati (Fig.~\ref{fig:hmap_guwahati}). As expected air traffic takes the infection to distant big cities while the roads restrict the infection in geographical proximity. When all data is combined, the map (Fig.~\ref{fig:hmap_bangalore}d, Fig.~\ref{fig:hmap_guwahati}d) is largely influenced by  rail and road traffic patterns due to their higher contribution to the total traffic.

Earlier works which used mobility to study the spread have exclusively used airline mobility, which is justified in the global context \cite{Science-Deff}. India has not just one of the largest railway networks but is also used by a significant fraction of people. Hence, an analysis of this type presented here is most desirable in the Indian context.

\begin{table}[!h]
\centering
\begin{tabular}{|l|c|c|}
\hline
\multicolumn{3}{|c|}{Delhi}              \\ \hline
\multicolumn{1}{|c|}{City} & \Deff & TOA  \\ \hline
Kanpur                     & 3.96 & 1.62 \\ \hline
Mumbai                     & 4.06 & 1.69 \\ \hline
Gurgaon                    & 4.25 & 1.94 \\ \hline
Lucknow                    & 4.33 & 2.00    \\ \hline
Faridabad                  & 4.34 & 2.00   \\ \hline
Jhansi                     & 4.54 & 2.19 \\ \hline
Rohtak                     & 4.58 & 2.31 \\ \hline
Ludhiana                   & 4.70  & 2.38 \\ \hline
Moradabad                  & 4.70  & 2.44 \\ \hline
Bangalore                  & 4.71 & 2.38 \\ \hline
\end{tabular} \quad \quad\quad \begin{tabular}{|l|c|c|}
\hline
\multicolumn{3}{|c|}{Mumbai}                                                      \\ \hline
\multicolumn{1}{|c|}{City} & \multicolumn{1}{c|}{\Deff} & \multicolumn{1}{c|}{TOA} \\ \hline
Thane                      & 2.89                      & 1.00                        \\ \hline
Pune                       & 3.17                      & 1.19                     \\ \hline
Delhi                      & 3.7                       & 1.62                     \\ \hline
Surat                      & 4.07                      & 2.00                        \\ \hline
Ahmedabad                  & 4.08                      & 2.00                       \\ \hline
Pimpri Chinchwad           & 4.25                      & 2.19                     \\ \hline
Nashik                     & 4.33                      & 2.25                     \\ \hline
Vasai                      & 4.43                      & 2.38                     \\ \hline
Vasco Da Gama              & 4.47                      & 3.00                       \\ \hline
Bangalore                  & 4.49                      & 2.38                     \\ \hline
\end{tabular}

\vspace{3em}

\begin{tabular}{|l|c|c|}
\hline
\multicolumn{3}{|c|}{Patna}                                                       \\ \hline
\multicolumn{1}{|c|}{City} & \multicolumn{1}{c|}{\Deff} & \multicolumn{1}{c|}{TOA} \\ \hline
Gaya                       & 2.98                      & 2.06                     \\ \hline
Dinapur Nizamat            & 3.32                      & 2.50                      \\ \hline
Arrah                      & 3.58                      & 2.75                     \\ \hline
Delhi                      & 3.75                      & 2.81                     \\ \hline
Bhagalpur                  & 3.99                      & 3.25                     \\ \hline
Kolkata                    & 4.09                      & 3.25                     \\ \hline
Darbhanga                  & 4.44                      & 3.88                     \\ \hline
Jehanabad                  & 4.47                      & 3.94                     \\ \hline
Begusarai                  & 4.57                      & 4.00                        \\ \hline
Biharsharif                & 4.60                       & 3.94                     \\ \hline
\end{tabular}\quad \quad\quad\begin{tabular}{|l|c|c|}
\hline
\multicolumn{3}{|c|}{Tirupati}                                                    \\ \hline
\multicolumn{1}{|c|}{City} & \multicolumn{1}{c|}{\Deff} & \multicolumn{1}{c|}{TOA} \\ \hline
Chittoor                   & 2.41                      & 2.88                     \\ \hline
Chennai                    & 2.53                      & 2.88                     \\ \hline
Hyderabad                  & 3.04                      & 3.50                      \\ \hline
Vellore                    & 4.23                      & 5.31                     \\ \hline
Bangalore                  & 4.25                      & 5.06                     \\ \hline
Tiruvannamalai             & 4.50                       & 5.81                     \\ \hline
Kadapa                     & 4.84                      & 6.50                      \\ \hline
Vijayawada                 & 4.89                      & 6.62                     \\ \hline
Anantapur                  & 5.00                         & 6.75                     \\ \hline
Madanapalle                & 5.01                      & 6.75                     \\ \hline
\end{tabular}
\caption{Time of Arrival $(T_A)$, in days, for each of the four outbreak locations in 
Fig. \ref{fig:deff_vs_dgeo}, showing cities with largest 10 values of $T_A$. 
The parameters are $\alpha=1.5, \beta=1.0,$ and $I^c=10$.}
\label{table:4cities_hi}
\end{table}

Finally, in Table \eqref{table:real-life}, the results of our framework are compared with real data of $T_A$ for the first wave of the SARS-CoV-2 pandemic. 

District-wise data is available from 26th April, 2020 \cite{covid19india}. Mumbai crossed the threshold first and is taken to be the outbreak location. There were $\sim 4000$ active cases in Mumbai on 26th April 2020. The $T_A$ at a city is when its three-day average caseload crosses a threshold taken to be $I^c=4000$. 
Table \eqref{table:real-life} presents a comparison of the real data with predictions from $D_{\rm eff}$. 
Table \eqref{table:real-life} (left) has the top 12 at-most risk cities based on the \Deff framework. This is compared with their rank in terms of real-life time of arrival of infections $T_A$. We find that 9 out the 12 cities also appear in the top-12 based on estimates from $T_A$. To present a different means of comparison, Table \eqref{table:real-life} (right) has the top-12 cities based on time of arrival. Again we find that 9 out of these 12 appear in the top-12 based on \Deff. It is remarkable, given the uncertainties in the traffic data and the approximations made to fill in missing data, that $\sim$ 75\% of cities obtained from simulations match with those in the list obtained from real data
This provides the proof-of-concept that it is possible to create a systematic predictive framework to objectively estimate the risk in Indian cities \cite{website}.

\begin{table}[!h]
\centering
\begin{tabular}{|c|l|c|c|c|c|l|c|c|}
\cline{1-4} \cline{6-9}
\begin{tabular}[c]{@{}c@{}}Rank \\ based \\ on \Deff\end{tabular} &
  \multicolumn{1}{c|}{City} &
  \Deff &
  \begin{tabular}[c]{@{}c@{}}Rank \\ based \\ on $T_A$\end{tabular} &
  \multirow{13}{*}{} &
  \begin{tabular}[c]{@{}c@{}}Rank \\ based \\ on $T_A$\end{tabular} &
  \multicolumn{1}{c|}{City} &
  \begin{tabular}[c]{@{}c@{}}$T_A$\\  (in days)\end{tabular} &
  \begin{tabular}[c]{@{}c@{}}Rank \\ based \\ on \Deff\end{tabular} \\ \cline{1-4} \cline{6-9}  
1  & \textbf{Thane}     & \textbf{2.88} & \textbf{4}  &  & 1  & \textbf{Delhi}     & \textbf{11} & \textbf{3}  \\ \cline{1-4} \cline{6-9} 
2  & \textbf{Pune}      & \textbf{3.18} & \textbf{5}  &  & 2  & \textbf{Ahmedabad} & \textbf{13} & \textbf{5}  \\ \cline{1-4} \cline{6-9} 
3  & \textbf{Delhi}     & \textbf{3.70} & \textbf{1}  &  & 3  & \textbf{Chennai}   & \textbf{16} & \textbf{12} \\ \cline{1-4} \cline{6-9} 
4  & Surat              & 4.06          & 222         &  & 4  & \textbf{Thane}     & \textbf{25} & \textbf{1}  \\ \cline{1-4} \cline{6-9} 
5  & \textbf{Ahmedabad} & \textbf{4.08} & \textbf{2}  &  & 5  & \textbf{Pune}      & \textbf{46} & \textbf{2}  \\ \cline{1-4} \cline{6-9} 
6  & \textbf{Nashik}    & \textbf{4.29} & \textbf{10} &  & 6  & \textbf{Hyderabad} & \textbf{57} & \textbf{10} \\ \cline{1-4} \cline{6-9} 
7  & Vasai              & 4.42          & No Data     &  & 7  & \textbf{Bangalore} & \textbf{65} & \textbf{9}  \\ \cline{1-4} \cline{6-9} 
8  & Vasco Da Gama      & 4.47          & No Data     &  & 8  & Guwahati           & 70          & 55          \\ \cline{1-4} \cline{6-9} 
9  & \textbf{Bangalore} & \textbf{4.49} & \textbf{7}  &  & 9  & \textbf{Kolkata}   & \textbf{79} & \textbf{11} \\ \cline{1-4} \cline{6-9} 
10 & \textbf{Hyderabad} & \textbf{4.62} & \textbf{6}  &  & 10 & \textbf{Nashik}    & \textbf{85} & \textbf{6}  \\ \cline{1-4} \cline{6-9} 
11 & \textbf{Kolkata}   & \textbf{4.91} & \textbf{9}  &  & 11 & Guntur             & 88          & 141         \\ \cline{1-4} \cline{6-9} 
12 & \textbf{Chennai}   & \textbf{4.93} & \textbf{3}  &  & 12 & Kurnool            & 89          & 131         \\ \cline{1-4} \cline{6-9} 
\end{tabular}
\caption{Comparison of the time of arrival of infection for real data and that from the simulation framework proposed in this work. The \textit{Time of Arrival} $T_A$ is given in days. The table shows top-12 at most risk cities obtained through \Deff and real-life data. The ranks based on $T_A$ for top-12 cities according to \Deff are given. Correspondingly, a list is prepared for top-12 cities based on $T_A$. The rows in bold denote the cities common to both lists. Note that 9 out of 12 cities ($\sim$ 75 \%) are common to both the lists, showing that the proposed framework has predictive capability. Mumbai is taken to be the outbreak location and the real-life data is at the granularity of districts.}
\label{table:real-life}
\end{table}

\section{Conclusion}
\label{chap:conclusion}

If an infectious disease breaks out in one city, how long does it take to reach other cities and towns? This length of time can be a simple measure of the risk in other cities -- the longer it takes, the lesser the risk. One may estimate this from careful simulations involving detailed traffic patterns. However this time is easily predicted by a quantity called the effective distance, which can be calculated if we know the prevailing traffic patterns. Larger the effective distance of a city from an outbreak location, lower is its risk of early infection.

Based on this idea, we have constructed an infectious disease hazard map for India using the data from the inter-city transportation network in India. Further details about the map can be found at \url{https://www.iiserpune.ac.in/~hazardmap/}.

Real data from air, road, and rail transportation networks between the most populous 446 Indian cities was used in this calculation. We relied on publicly available data sources and used simple assumptions and algorithms to fill-in the missing attributes of typical Indian traffic patterns.

We used extensive simulations to validate the usefulness of the idea. We find, in agreement with similar past studies, that the effective distance of a city from the origin is proportional to the time of appearance of first infections in that city and is thus a reliable measure of its risk. Comparison with the early patterns of spread of COVID-19 in India showed surprisingly good agreement between the predictions from effective distances and real data. This adds further credence to the idea of effective distance.

The results here prompt several interesting questions both of academic and practical value. While effective distance predicts relative order in which cities are affected, the rate of spread through this sequence is determined by the details of the infection parameters ($\alpha,\beta$) as well as average mobility. A conceptual framework that explains the empirical observations regarding these (from simulations) is missing. Moreover a good explanation for the linear relationship between the effective distance and time of arrival may be needed in order to know the limits of its applicability. Lastly, a generalization of the notion of effective distance to a scenario of multiple outbreak locations will make this an invaluable tool in designing efficient mitigation measures -- for instance in determining which traffic routes to close down with higher priority.

\section*{Acknowledgments}
This work was funded by a special MATRICS grant MSC/2020/000122 given by SERB, Government of India to MSS, GJS and SJ. OS would like to thank Aanjaneya Kumar and Suman Kulkarni for the discussions. OS and MB acknowledge the INSPIRE grant from the Department of Science and Technology, India. Map plots in the paper are made using Leaflet \cite{map}


\begin{thebibliography}{99}

\bibitem{WHO-Hopkins}{\url{https://covid19.who.int/}, \url{https://coronavirus.jhu.edu/map.html} }

\bibitem{covid19india}{\url{https://www.covid19india.org/}}


\bibitem{mills} I. D. Mills, The 1918-1919 influenza pandemic - The Indian experience, The Indian Economic and 
Social History Review {\bf 23}, 1 (1986). {\url{https://doi.org/10.1177\%2F001946468602300102}}

\bibitem{sani-comm} The Annual Report of the Sanitary Commissioner with the Government of India (1918), (Calcutta, 1920).

\bibitem{arrival-time-statistics}{Gautreau, A., Barrat, A., Barthelemy, M., Global disease spread: statistics and estimation of arrival times. Journal of theoretical biology \textbf{251}, 509–522. (2008)}

\bibitem{Barthelemy-network}{Barthélemy, M., Spatial networks. Physics Reports \textbf{499}, 1–101. (2011)}

\bibitem{Epidemic-networked}{Feng, L., Zhao, Q., Zhou, C., Epidemic in networked population with recurrent mobility pattern. Chaos, Solitons \& Fractals 139, 110016. (2020)}



\bibitem{Mobility-infection}{Belik, V., Geisel, T., Brockmann, D., Natural human mobility patterns and spatial spread of infectious diseases. Physical Review X \textbf{1}, 011001. (2011)}

\bibitem{Global-Travel-threats}{Coltart, C.E., Behrens, R.H., The new health threats of exotic and global travel. British Journal of General Practice \textbf{62}, 512–513. (2012)}

\bibitem{Helbing-global-travel-risks}{Helbing, D., Globally networked risks and how to respond. Nature \textbf{497}, 51–59. (2013)}

\bibitem{Barthelemy-mobility}{Barbosa, H., Barthelemy, M., Ghoshal, G., James, C.R., Lenormand, M., Louail, T., Menezes, R., Ramasco, J.J., Simini, F., Tomasini, M., 2018. Human mobility: Models and applications. Physics Reports \textbf{734}, 1–74. (2018)}

\bibitem{Science-Deff}{Brockmann, D., Helbing, D., The Hidden Geometry of Complex, Network-Driven Contagion Phenomena. Science \textbf{342}, 1337–1342. (2013)} 

\bibitem{Covid19-mobility}{Arenas, A., Cota, W., Gómez-Gardeñes, J., Gómez, S., Granell, C., Matamalas, J.T., Soriano-Paños, D., Steinegger, B., Modeling the Spatiotemporal Epidemic Spreading of COVID-19 and the Impact of Mobility and Social Distancing Interventions. Physical Review X \textbf{10}, 041055. (2020)}

\bibitem{Covid19}{Althouse, B.M., Wenger, E.A., Miller, J.C., Scarpino, S.V., Allard, A., Hébert-Dufresne, L., Hu, H., Stochasticity and heterogeneity in the transmission dynamics of SARS-CoV-2. {\tt{arXiv}:2005.13689}. (2020)}

\bibitem{Global-DS-Covid}{Garcia-Gasulla, D., Napagao, S.A., Li, I., Maruyama, H., Kanezashi, H., P’erez-Arnal, R., Miyoshi, K., Ishii, E., Suzuki, K., Shiba, S.,  Global Data Science Project for COVID-19 Summary Report. {\tt{arXiv:2006.05573.}} (2020)}


\bibitem{India-Covid5}{Mishra, R., Gupta, H.P., Dutta, T., Analysis, Modeling, and Representation of COVID-19 Spread: A Case Study on India, in IEEE Transactions on Computational Social Systems, doi: 10.1109/TCSS.2021.3077701. (2021)}


\bibitem{India-Covid6}{Sarkar, K., Khajanchi, S., Nieto, J.J., Modeling and forecasting the COVID-19 pandemic in India. Chaos, Solitons \& Fractals \textbf{139}, 110049. (2020)}

\bibitem{SPPU-India_Covid}{Pujari, B.S., Shekatkar, S., Multi-city modeling of epidemics using spatial networks: Application to 2019-nCov (COVID-19) coronavirus in India. medRxiv. \url{https://doi.org/10.1101/2020.03.13.20035386} (2020)}

\bibitem{IITB-India_Covid}{Gupta, S., Shah, S., Chaturvedi, S., Thakkar, P., Solanki, P., Dibyachintan, S., Roy, S., Sushma, M.B., Godbole, A., Jaseem, N. \textit{et. al.}, An India-specific compartmental model for Covid-19: projections and intervention strategies by incorporating geographical, Infrastructural and response heterogeneity. {\tt{arXiv:2007.14392}}. (2020)}


\bibitem{India-Covid0} R. Gopal, V. K. Chandrasekar and M. Lakshmanan,  Dynamical modelling and analysis of  COVID-19 in India, Current Science {\bf 120} 1342 (2021).

\bibitem{India-Covid1}{Bedi, P., Dhiman, S., Gole, P., Jindal, V., Prediction of COVID-19 Trend in India and Its Four Worst-Affected States Using Modified SEIRD and LSTM Models. SN COMPUT. SCI. \textbf{2}, 224 (2021)}

\bibitem{India-Covid2}{Khajanchi, S., Sarkar, K., Forecasting the daily and cumulative number of cases for the COVID-19 pandemic in India. Chaos: An Interdisciplinary Journal of Nonlinear Science \textbf{30}, 071101. (2020)}

\bibitem{ICTS-India_Covid}{Das, A., Dhar, A., Goyal, S., Kundu, A., Pandey, S., COVID-19: Analytic results for a modified SEIR model and comparison of different intervention strategies. Chaos, Solitons \& Fractals 110595. (2021)}

\bibitem{India-Covid3}{Jha, V., Forecasting the transmission of Covid-19 in India using a data driven SEIRD model. {\tt{arXiv:2006.04464.}} (2020)}

\bibitem{India-Covid4}{Khajanchi, S., Sarkar, K., Mondal, J., Perc, M., Dynamics of the COVID-19 pandemic in India. {\tt{arXiv:2005.06286.}} (2020)}




\bibitem{census}{\url{https://censusindia.gov.in/2011-common/censusdata2011.html}}


\bibitem{SIR-meta1} {Gong, Y., Song, Y., Jiang, G.,
Epidemic spreading in metapopulation networks with heterogeneous infection rates, Physica A: Statistical Mechanics and its Applications,
\textbf{416}, 208-218. (2014)}

\bibitem{SIR-1}{Ronald, R., and Hilda, P., An application of the theory of probabilities to the study of a priori pathometry.--Part IIIProc. R. Soc. Lond. A \textbf{93}: 225–240. (1917)}

\bibitem{SIR-2}{Kermack, W. O. and McKendrick, A. G., A contribution to the mathematical theory of epidemics, Proc. R. Soc. Lond. A \textbf{115}:700–721. (1927)}

\bibitem{SIR-meta2}{Colizza, V., Pastor-Satorras, R., and Vespignani, A. Reaction-diffusion processes and metapopulation models in heterogeneous networks. Nature Phys \textbf{3}, 276–282 (2007).}

\bibitem{TDA-spreading}{Taylor, D., Klimm, F., Harrington, H.A., Kramár, M., Mischaikow, K., Porter, M.A., Mucha, P.J., 2015. Topological data analysis of contagion maps for examining spreading processes on networks. Nature communications \textbf{6}, 1–11. (2015)}

\bibitem{Shortest-path-SIR}{Tolić, D., Kleineberg, K.-K., Antulov-Fantulin, N., Simulating SIR processes on networks using weighted shortest paths. Scientific reports \textbf{8}, 1–10. (2018)}




\bibitem{supplement}{Please refer to the Supplementary material.}

\bibitem{r0info}{Marimuthu S., Joy M.,Malavika  B., Nadaraj A., Asirvatham E., Jeyaseelan L., Modelling of reproduction number for COVID-19 in India and high incidence states. Clinical Epidemiology and Global Health, \textbf{9}, 57-61. (2021)}

\bibitem{website}{For more information, visit \url{https://www.iiserpune.ac.in/~hazardmap/}}

\bibitem{map} \href{https://leafletjs.com/}{Leaflet} | Map tiles by \href{https://stamen.com/}{Stamen Design}, under \href{https://creativecommons.org/licenses/by/3.0/}{CC BY 3.0.} Data by © \href{https://www.openstreetmap.org/#map=6/21.850/82.675}{OpenStreetMap}, under \href{https://creativecommons.org/licenses/by-sa/3.0/}{CC BY SA}.



\end{thebibliography}
\end{document}


\maketitle
\section{Transportation network and data}
\label{chap:data}
In this supplementary material, network creation and data sources used in estimating the traffic matrix ${\mathbf F}$ are discussed. For the purposes of this work, we focus on data from air, railway and road transportation while ignoring inland waterways and other modes.

Based on the 2011 census data \cite{census}, a network of cities/towns in India whose population is more than 1 Lakh and having at least one of air, rail or road connectivity is created. By this criterion, a directed network with $M=446$ nodes (cities/towns), and 46448 edges (routes) is created. In this, the edges corresponding to air, railway and road links are combined if they connect the same pair of cities and are counted as one edge. Some  important network properties are listed in Table \eqref{table:trafficdata}. In the table, mean degree refers to the out-degree (which is the same as in-degree for this dataset) and gives a measure of the average number of cities each city is connected to.  This table also shows coarse mobility data. By `route symmetry index' in the context of pair of cities $a$ and $b$, it is indicated if route $a \to b$ has the same traffic as the reverse route $b \to a$ or not. The index is defined for a given dataset as `$1-$ the maximum absolute fractional traffic difference with respect to the city's population averaged over all cities'. The index can mathematically be expressed as
\begin{align}
\text{Route Symmetry Index} &=1-\left\langle \frac{\text{max}_j \{ |F_i^j-F_j^i |\}}{N_i} \right\rangle_i.
\end{align}  
The locality of mobility indicates if the fraction of people who travel out of a city, i.e. $\frac{F_i}{N_i}$, is the same across all cities or not. As is evident, air travel contributes far less to the overall mobility of people in India though it might become somewhat important if the speed of transportation is taken into account. As road travel is a locally ($<$ 300 km) dominant mode, most of the meaningful long-distance 
connections and mobility arise from railways.

\begin{table}[!h]
\centering
\renewcommand{\arraystretch}{1.5}
\begin{tabular}{|l|c|c|c|c|}
\hline
Property & \multicolumn{1}{l|}{Airway} & \multicolumn{1}{l|}{Railway} & \multicolumn{1}{l|}{Roadway} & \multicolumn{1}{l|}{Combined} \\ [2.0mm]\hline 
Number of Nodes & 85 & 435 & 446 & 446 \\ [2.0mm]\hline
Number of Edges & 1182 & 41594 & 9128 & 46448 \\ [2.0mm]\hline
Average Degree & 13 & 95 & 20 & 104 \\ [2.0mm]\hline
Route symmetry index & 1 & 0.9875 & 1 & 0.9878 \\ [2.0mm]\hline
Locality of Mobility & Same & Different & Same & Different \\ [2.0mm]\hline
Passengers/day & $7.5 \times 10^5$ & $8.8 \times 10^6$ & $2.5 \times 10^6$ & $1.2 \times 10^7$ \\ [2.0mm]\hline
Fraction of total & 0.06 & 0.73 & 0.21 & 1.0 \\ [2.0mm]\hline
\end{tabular}
\caption{Properties of the transportation network and mobility data assembled for this work. Not surprisingly,
air travel constitutes a miniscule fraction of the overall mobility. Bulk of long distance travel is accounted
for by trains, though at short distances roads might also have significant contribution.}
\label{table:trafficdata}
\end{table}

{\it Railway data} : The data of trains and railway stations was obtained from the Indian Railways website, and travel planning websites \cite{train-station-data1,train-station-data2}. In all, Indian railway network has about 8000 railway stations through which about 3000 express/mail/passenger trains are run daily \cite{train-schedule}. The weekly schedule of each of these trains is drawn from the railway
timetable \cite{train-schedule}. Even though there are studies conducted to analyse the Indian railway network \cite{Indian-railway, Train-Capacity-India, Indian-railway2}, none of them considered the number of passengers travelling between two cities. The data of the number of trains run is converted into passenger traffic through an algorithm described in the next sections. The algorithm relies on two simple assumptions -- that the train is always full (almost always true in India), and the number of passengers travelling between two cities on some route is proportional to the population of the two cities. Based on this algorithm, the daily passengers on the train comes to about 88 Lakh [Table \eqref{table:trafficdata}]. As we are only concerned with long-distance traffic, the suburban traffic in several cities such as Mumbai and Chennai, which have a dense suburban train network, is also ignored.

{\it Airline data} : The passenger data on domestic routes, based on the data available on the Directorate General of Civil Aviation, Govt. of India (DGCA), was obtained \cite{air-data1} for 85 Indian cities with an airport and with a population of 1 Lakh or more. Passenger traffic was collected for pairs of cities for the year 2018, from which daily traffic was arrived at. According to the DGCA, the total domestic air traffic in India for 2018-2019 was 2851 Lakhs \cite{air-data2}. This implies that about 7.8 Lakhs people use air transport 
daily on an average, which upon correcting for the ignored airports for small cities with a population less than 1 Lakh, comes
to approximately 7.5 Lakhs per day, the value indicated in Table \eqref{table:trafficdata}.

{\it Road data} : Compared to the other two modes of transport, very little reliable information is available for road transport in India. The National Highways Authority of India (NHAI) provides
some minimal data only for national highways, which typically carry traffic over long-distances of 
$>$300-400 Kms. However, we ignore the long-distance road traffic by assuming that people mostly prefer train or air for long-distance travel. Further, the significant passenger flux over short distances of about $<$400 km is mainly accounted for by public transport and in some cases by private vehicles. Since
there is no reliable data available to estimate the short distance ($<$400 Kms) travel (or even the long-distance traffic) between any two given cities, we estimate these numbers using our algorithm given in
the next section. The basic idea behind the algorithm is that most people prefer road for distances of $<$ 300 Kms. This algorithm gives us daily traffic of around 25 Lakhs per day.


All these data are assembled together to obtain the averaged daily traffic matrix ${\mathbf F}$, whose
element $F_i^j$ represents the mean composite volume of traffic (number of people) in the route from 
city $i$ to $j$ 
in a ``typical" day. Note that the data is mainly drawn from or estimated for pre-covid 
years, 2017 to 2019. Based on
this dataset, we find that $F_i^j\neq F_j^i$, {\it i.e.}, it is a directed network, implying that the number of people travelling from city $i$ to $j$ and vice-versa are not equal. In this paper, we will also neglect the time-scales  associated with various modes of transport (e.g., the air is faster than train or road travel) and focus only on the average number of people travelling in a day.

\section{Algorithms}

The raw data that we have collected is not in the format of number of passengers travelling from city $i$ to city $j$ except for the airway. The rail data is in the form of travel routes, while there is no reliable source for road data. In this supplemental material, we will look at two algorithms to estimate the road and rail traffic matrices entries by making some reasonable assumptions.

\subsection{Railways}\label{app:rail_algo}
The raw data for railways is in the format of the train routes with stops \cite{train-station-data1, train-station-data2}. We have the population for each of the stop on all routes. The raw data set consists of 4480 routes, including both forward and backward routes \cite{train-schedule}. To make the data-analysis tractable, we include only those cities that have a population of more than 1 Lakh as per the 2011 census in India \cite{census}. Thus, few smaller sub-stations in the vicinity of major metropolitan cities are also treated as separate cities. 

To extract data from train routes and population of cities, we rely on the two assumptions,
\begin{enumerate}
    \item The train is always full while travelling between any two consecutive cities.
    \item The number of people travelling between any two cities (not necessarily consecutive) on a route is proportional to their individual populations.
\end{enumerate}

In addition to the information about the route and population, we also have the information about the weekly frequency and the type of train on each of the routes. We use the knowledge about the train type to fix the capacity of each train. We will now delve into the details of the algorithm.

Let us consider a particular route $\Omega$ with $k$ stations labelled by indices $1,\;2,\;...,\;k$. The population of the cities is $N_1,\; N_2,\; ...,\;N_k$ respectively. Assumption 1 tells us that the train starts with full capacity from city 1, which we denote by $C$. According to the assumption 2 these $C$ number of people will get down at remaining stations proportional to the population. Going back to our first assumption; the number of people getting on the train at any city $j$ ($j \geq 2$) would be the same as number of people getting down at that station. Let \textbf{F} denote the traffic matrix, where $F_i^j$ corresponds to number of people going from city $i$ to city $j$ in unit time. Similarly, $F_i$ represents the total number of people moving out of city $i$ in a unit amount of time.

Now, the number of people getting on the train at city 2, will again get down at cities $j$ ($j \geq 3)$, proportional to their population. The number of people getting down at city 3 would be dependent on number of people who got on the train in city 1 and 2. Thus, we continue this process until city $k-1$, and finally $C$ number of people will get down at city $k$. The total outgoing/incoming number of people for the cities on the route $\Omega$ can be expressed mathematically as,

\begin{align}
    F_1 (\Omega)&= C, \cr
    F_2 (\Omega)&= F_1 \; \frac{N_2}{\sum_{j=2}^k N_j},  \cr
    F_3 (\Omega)&= F_1 \; \frac{N_3}{\sum_{j=2}^k N_j} + F_2 \; \frac{N_3}{\sum_{j=3}^k N_j},\cr
    F_4 (\Omega)&= F_1 \; \frac{N_4}{\sum_{j=2}^k N_j} + F_2 \; \frac{N_4}{\sum_{j=3}^k N_j} + F_3 \;  \frac{N_4}{\sum_{j=4}^k N_j},\cr
    &\vdotswithin{=} \cr
    F_a (\Omega)&= \sum_{i=1}^{a-1}\Bigg[ F_i \;  \frac{N_a}{\sum_{j=i+1}^k N_j} \Bigg].
\end{align}

We note that these equations have a recursive form. Luckily, we can simplify each of them to get a simple form for $F_a$, which denotes the total number of people leaving city $a$ for route $\Omega$,

\begin{align}
    F_a (\Omega)&= F_1 \; \frac{N_a}{\sum_{j=a}^k N_j} = C \; \frac{N_a}{\sum_{j=a}^k N_j}, \qquad\qquad  \text{for,} \quad 2\leq a \leq k-1, \cr
    F_a (\Omega)&= F_1 = C,  \qquad \qquad \qquad \qquad \qquad \qquad \quad \; \text{for,} \quad a=1.
\end{align}

For instance the expression for $F_3 (\Omega)$ can be simplified as,

\begin{align}
    F_3 (\Omega)&= F_1 \; \frac{N_3}{\sum_{j=2}^k N_j} + F_2 \;\frac{N_3}{\sum_{j=3}^k N_j},\cr
    F_3 (\Omega)&= F_1 \; \frac{N_3}{\sum_{j=2}^k N_j} + F_1 \;\frac{N_2}{\sum_{j=2}^k N_j} \;\frac{N_3}{\sum_{j=3}^k N_j},\cr
    F_3 (\Omega)&= F_1 \;\frac{N_3}{\sum_{j=2}^k N_j} \Bigg[1+\frac{N_2 }{\sum_{j=3}^k N_j} \Bigg], \cr
    F_3 (\Omega)&= F_1 \;\frac{N_3}{\sum_{j=2}^k N_j} \Bigg[\frac{\sum_{j=2}^k N_j}{\sum_{j=3}^k N_j} \Bigg],\cr
    F_3 (\Omega)&= F_1 \; \frac{N_3}{\sum_{j=3}^k N_j}.
\end{align}

The number of people going from city $a$ to city $b$ ($2 \leq a < b \leq k$) for the route $\Omega$ would be,

\begin{align}
    F_a^b (\Omega)&= \Bigg[ C \;\frac{N_a}{\sum_{j=a}^k N_j} \Bigg] \Bigg[\frac{N_b}{\sum_{j=a+1}^k N_j} \Bigg], \; \qquad \qquad  \text{for,} \quad 2\leq a <b \leq k, \cr
        F_a^b (\Omega)&= \Bigg[ C \frac{N_b}{\sum_{j=a+1}^k N_j}  \Bigg], \qquad \qquad \qquad \qquad \quad \quad  \text{for,} \quad 1=a < b \leq k.\label{eq:algo_train_Fij}
\end{align}

\begin{figure}[!h]
    \centering
    \includegraphics[width=\columnwidth]{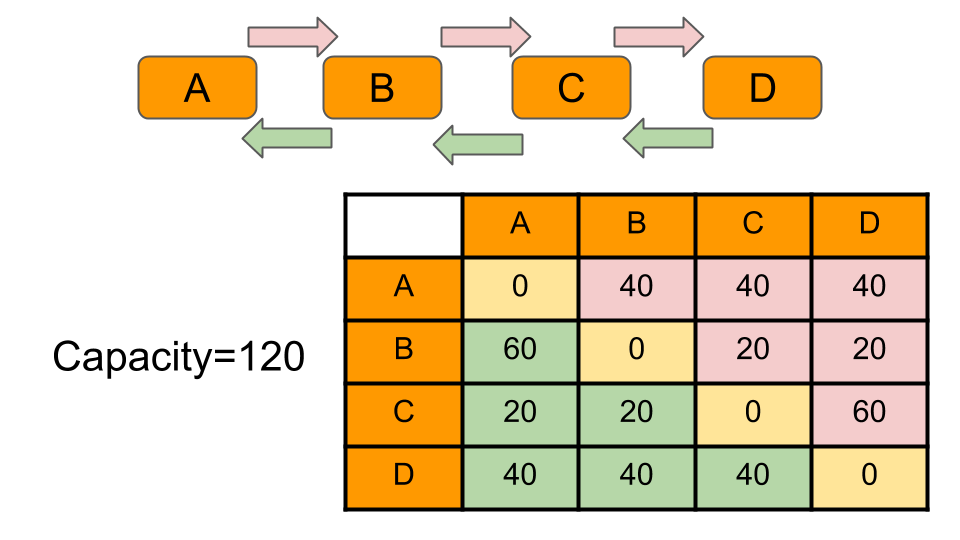}
    \caption{Model Train Algorithm: We consider two routes here: A-B-C-D and D-C-B-A. For simplicity let us assume that the population of all four cities is equal and the capacity of the train is 120. We run the algorithm using Eq.~\eqref{eq:algo_train_Fij} and each entry in the table $T_{ij}$ specifies the number of people travelling from $i$ to $j$. The colour in the table correspond to the route in the upper part of the figure.}
    \label{fig:train_algo}
\end{figure}

Thus, we have obtained the expression for number of people travelling from city $a$ to city $b$ for a given route. There is only one free parameter in the above expression $F_1= C$, which we fix by knowing the type of train and its capacity \cite{train-station-data1}. We can run the same algorithm for all the routes in our raw data-set to get the full \textbf{F}-matrix for railway as the mode of transport. We illustrate this algorithm in a more visual and simple way, using an imaginary toy route and population in Fig.~\eqref{fig:train_algo}.

Let us consider a train route going through 4 cities A, B, C, and D having equal population $N$. We will also assume that the capacity of train is 120. Figure.~\eqref{fig:train_algo} shows the final results after running the algorithm for the forward and backward routes. As we can readily see the \textbf{F}-matrix is not symmetric even if we consider the simplest case of just one route through cities of equal population. It is also worth noting that the total number of people travelling in/out of a city is not proportional to the city population. The assymetricity of \textbf{F}-matrix and unequal local mobility becomes more pronounced when we consider all cities and all the routes. 

\subsection{Road}\label{app:road_algo}
In this section, we will look at the algorithm used to generate the road traffic. To our knowledge, there is no systematic source of road traffic connecting almost all cities in India with population above 1 Lakh. The only available data is about toll booths on national highways \cite{road-data1}. This data is not useful as it leaves out most of the short-distance routes. Thus, we have to estimate the traffic data by using the available information.

As a simplest case, we assume that the final \textbf{F}-matrix is symmetric. We can move away from this assumption, however, the process to make sure that the population of each city remains constant becomes more complex.

The next assumption we make is that most people use road travel for a short distance and long-distance travel is usually undertaken through railways or by aeroplane. And we make the final assumption that the total number of people travelling by road is proportional to the population of the city. We will later see how these specific assumptions help us fix the free parameters in our algorithms under the given set of constraints. 

First we construct a graph of the cities. We consider cities within a specified distance as adjacent in this graph and construct an adjacency matrix for this graph. The distances are calculated using the latitudes and longitudes of the cities through the \textit{geopy} library in python. Thus, the distances are not the lengths of the road between two cities, which in general would be more than the air distance. We then run an algorithm similar to the one we used for the train for this adjacency matrix.

\begin{figure}[!h]
    \centering
    \includegraphics[width=0.45\columnwidth]{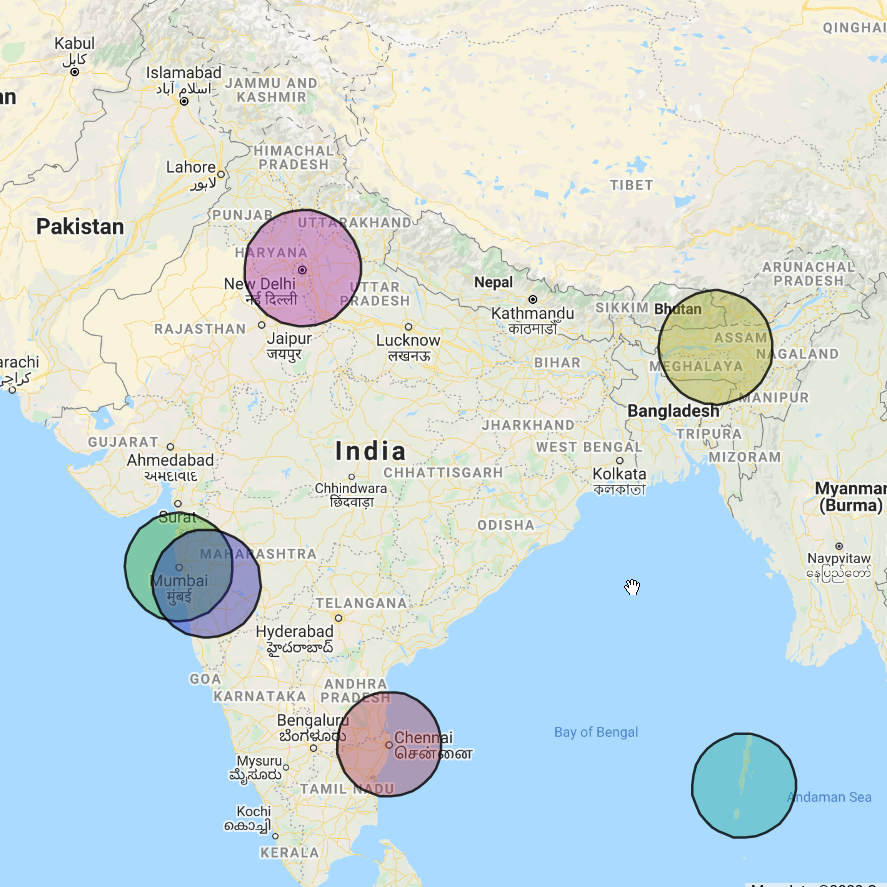}\hspace{0.05\columnwidth}\includegraphics[width=0.45\columnwidth]{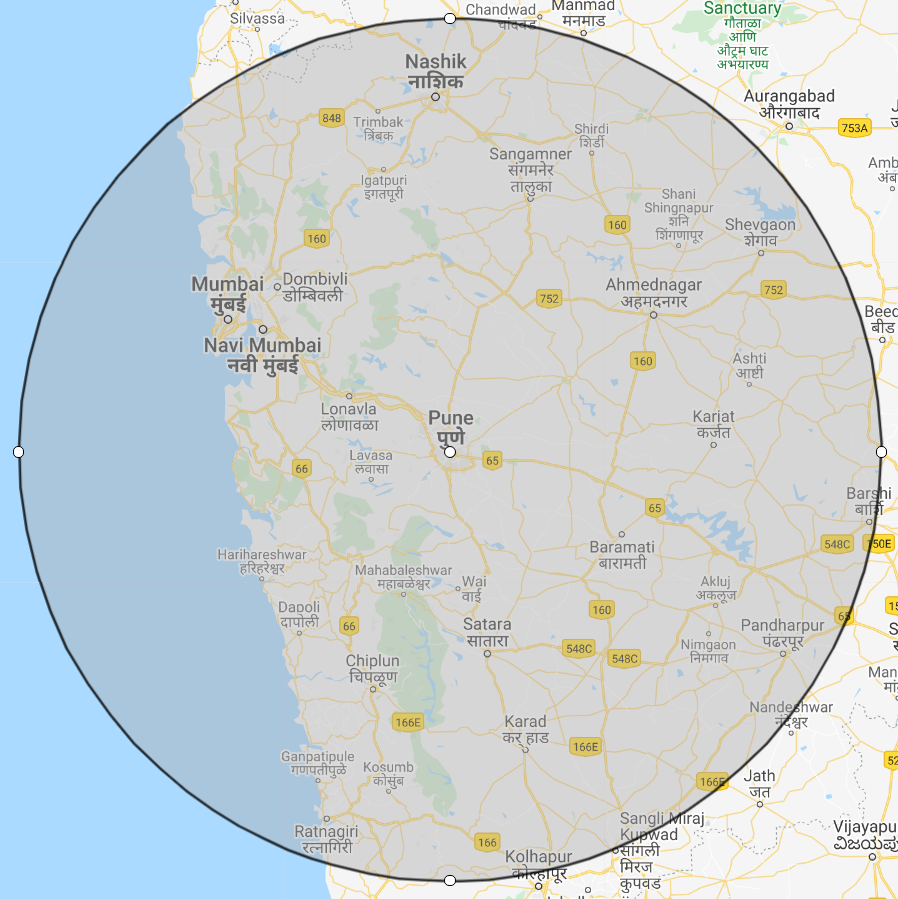}
    \caption{The relative size of 200 Kms circle for scale on map of India. On left hand side, we draw the circles around Mumbai, Pune, Delhi, Secennai, Guwahati, and Port-Blair. The right hand figure shows the circle for Pune. We can identify few of the important cities in Maharashtra which are connected to Pune by road from our algorithm.}
    \label{fig:road_algo_circles}
\end{figure}

Once we have the adjacency matrix, we sort the rows and columns according to the population. We start with the city with lowest population. We get the number of people travelling from this city by fixing a value for desired mobility and then distribute them to the cities adjacent to it proportional to their respective populations. We then make a symmetric entry in the \textbf{F}-matrix. Next we move on to the city with second lowest population. We get the number of people travelling by multiplying the city population with the desired mobility. We subtract those people who are already accounted for through the symmetric entry in the first step, and distribute the rest proportionally same as before. We repeat this process, until we reach the most populous city. In case, the number of people accounted due to symmetric entry is higher than the desired mobility, we simply increase the mobility for that particular city by a small amount.

\begin{figure}[h]
    \centering
    \includegraphics[width=0.8\columnwidth]{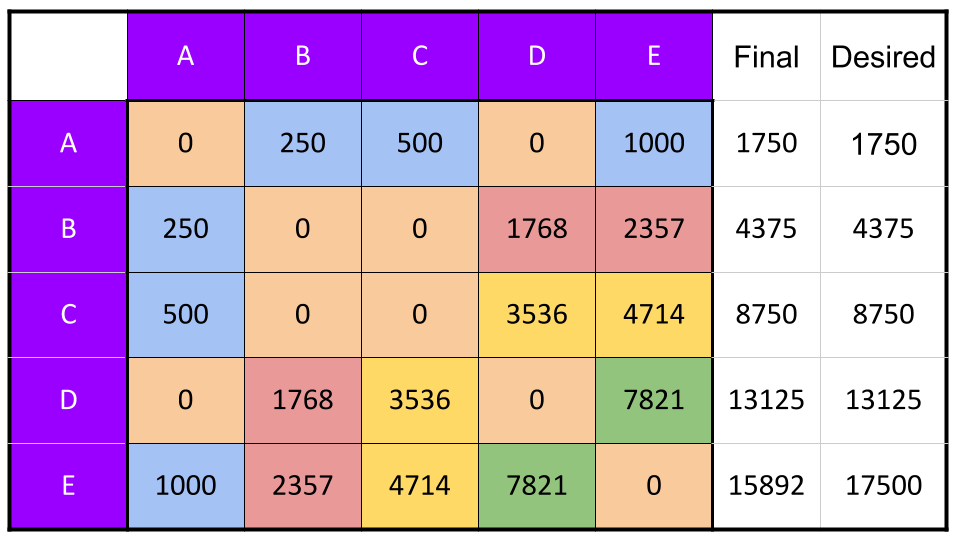}
    \caption{Road Algorithm: The \textbf{F}-matrix for the toy model is symmetric and the final traffic, matches exactly with desired traffic for most cities.}
    \label{fig:road_algo}
\end{figure}

In order to understand this algorithm more clearly, we again give a toy example to illustrate our case in Fig.~\eqref{fig:road_algo}. We consider a network of 5 cities, such that the links AD, DA, BC, and CB are missing. The population of the five cities are in proportion as follows: A : B : C : D : E :: 10 : 25 : 50 : 75 : 100. Let's suppose that the population of city A is 1 Lakh and desired mobility is 0.0175. The algorithm proceeds as follows: 

\begin{itemize}
    \item First we get the adjacency matrix whose indices are sorted according to the population. We start with lowest population city, i.e. A and distribute the desired number of people into all possible connections (blue coloured).
    \item We make symmetric entries in the first column (blue), subtract that number from desired traffic for city B and distribute the rest into possible connections (red). We make similar symmetric entry in red.
    \item For city C, we proceed as before, subtract, distribute, and make symmetric entry (yellow). 
    \item We continue the algorithm until the table is fully filled.
\end{itemize}

Coming back to our algorithm for a network of 446 cities, the parameter values that we used in our algorithm are,

\begin{itemize}
    \item desired mobility $\gamma$ = 0.015. 
    \item radius of circle of proximity = 200 Kms.
    \item The increase in traffic to distribute in case the symmetric entries already sum up more than the desired traffic was ($0.5 \times$ desired traffic). 
\end{itemize}

According to \cite{road-data2}, the total yearly road traffic in India is around 822 crore passengers. If we consider the average daily road traffic it comes out to be around 2.25 crore passengers. At a gross level, if we calculate the global mobility for India considering that the current population of 130 crores, we get the global mobility to be around $0.017$. We take it to be $0.015$ in order to compensate for fluctuations and other uncertainties.

We now mention few of the main results of the \textbf{F}-matrix for road transport.

\begin{enumerate}
    \item We increased the mobility of only 6 (out of 446) cities to account for overflow of traffic due to symmetric entries.
    \item The cities have 20 connections on an average.
    \item 92\% (or $\approx$ 410) cities have a local mobility of 0.015. The average (global) mobility is 0.0115 and the standard deviation of it is 0.0021.
\end{enumerate}
